\documentclass[11pt,a4paper]{article}

\pdfoutput=1 
\usepackage{ccaption}
\usepackage[utf8]{inputenc}
\usepackage{amsmath,amssymb}
\usepackage{mathtools}
\usepackage{setspace}
\usepackage{afterpage}
\usepackage{geometry}
\usepackage{graphicx}
\usepackage{sectsty}
\sectionfont{\fontsize{12}{15}\selectfont}
\usepackage[english]{babel}
\usepackage[nottoc]{tocbibind}
\usepackage{fancyhdr}
\usepackage{indentfirst}
\usepackage{multirow}
\usepackage{url}

\usepackage{soul} 
\usepackage{cite}
\usepackage{array}
\usepackage{color}
\definecolor{mygray}{gray}{0.6}

\newcolumntype{C}[1]{>{\centering\arraybackslash}p{#1}}

\newcommand{\beginsupplement}{%
        \setcounter{table}{0}
        \renewcommand{\thetable}{S\arabic{table}}%
        \setcounter{figure}{0}
        \renewcommand{\thefigure}{S\arabic{figure}}%
     }

\usepackage{float}
\floatstyle{plaintop}
\restylefloat{table} 
\usepackage{xcolor} 
\usepackage{graphicx}
\usepackage{lipsum}
\usepackage{ccaption}

\usepackage[normalem]{ulem} 

\usepackage[symbol]{footmisc}

\usepackage{verbatim}

\usepackage{hyperref}
\hypersetup{
}

\newgeometry{
    left=1.0in,
    right=1.0in,
    top=1in,
    bottom=1in
}

 \usepackage{afterpage} 
 \newcommand\blankpage{%
    \null
    \newpage}
 \usepackage{xr}
 \makeatletter
 \newcommand*{\addFileDependency}[1]{SI.tex
   \typeout{(#1)}
   \@addtofilelist{#1}
   \IfFileExists{#1}{}{\typeout{No file #1.}}
 }
 \makeatother
 
\title{Network physics of attractive colloidal gels: Resilience, Rigidity, and Phase Diagram}

\author{Mohammad Nabizadeh$^{1,*}$, \and Farzaneh Nasirian$^{2,3}$\footnote{Authors equally contributed.}, \and Xinzhi Li$^{3}$, \and Yug Saraswat$^{4}$, \and Rony Waheibi$^{4}$, \and Lilian C. Hsiao$^{4}$, \and Dapeng Bi$^{3}$, \and Babak Ravandi$^{2,3,\dagger}$, \and Safa Jamali$^{1,}$\footnote{Corresponding authors e-mail: bk.ravandi@gmail.com and s.jamali@northeastern.edu}
}

\date{%
    $^1${\small Department of Mechanical and Industrial Engineering, Northeastern University, Boston, MA, 02215, USA} 
    \vspace{-7pt}\\
    $^2${\small Network Science Institute, Northeastern University, Boston, MA, 02215, USA}
    \vspace{-7pt}\\
    $^3${\small Department of Physics, Northeastern University, Boston, MA, 02215, USA}
    $^{4}${\small Department of Chemical and Biomolecular Engineering, North Carolina State University, Raleigh, NC, 27606, USA}
    \vspace{-7pt}
 \\[2ex]%
}

\usepackage{lineno}

\begin{document}

\flushbottom
\maketitle

\section*{Abstract}
\vspace{-20pt}
\noindent\rule{6.3in}{0.4pt}
Attractive colloidal gels exhibit solid-like behavior at vanishingly small fractions of solids, owing to ramified space-spanning networks that form due to particle-particle interactions. These networks give the gel its rigidity, and as the attraction between the particles grows, so does the elasticity of the colloidal network formed. The emergence of this rigidity can be described through a mean field approach; nonetheless, fundamental understanding of how rigidity varies in gels of different attraction strengths is lacking. Moreover, recovering an accurate gelation phase diagram based on the system's variables have been an extremely challenging task. Understanding the nature of these fractal clusters, and how rigidity emerges from their connections is key to controlling and designing gels with desirable properties. Here, we employ well-established concepts of network science to interrogate and characterize the network of colloidal gels. We construct a particle-level network, having all the spatial coordinates of colloids with different attraction levels, and also identify polydisperse rigid fractal clusters using a Gaussian Mixture Model, to form a coarse-grained cluster network that distinctly shows main physical features of the colloidal gels. A simple mass-spring model then is used to recover quantitatively the elasticity of colloidal gels from these cluster networks. Interrogating the resilience of these gel networks show that the elasticity of a gel (a dynamic property) is directly correlated to its cluster network's resilience (a static measure). Finally, we use the resilience investigations to devise [and experimentally validate] a fully resolved phase diagram for colloidal gelation, with a clear solid-liquid phase boundary using a single volume fraction of particles well beyond this phase boundary.

\section*{Main}
\vspace{-20pt}
\noindent\rule{6.3in}{0.4pt}
Structure formation through self-aggregation of particles is ubiquitous in natural and industrial settings alike, with numerous examples from biological systems and living systems, to food processing and consumer products. In case of passive attractive colloids, this self-assembly at low and intermediate volume fractions of solid, results in space-spanning out-of-equilibrium structures \cite{zaccarelli2007colloidal}, that are commonly referred to as ``gels'' exhibiting a wide range of mechanical and rheological properties \cite{stradner2004equilibrium, poon2004colloids, anderson2002insights}. Particle-particle bonds formed due to attractive surface forces above a certain threshold of solid particles eventually construct particulate networks that in turn govern the mechanical and rheological properties of colloidal gels. Of particular interest has been emergence of rigidity in relatively small solid fraction, and a phase transition from liquid- or even gas-like to solid-like behavior \cite{krall1998internal, hsiao2012role, lu2008gelation, trappe2001jamming, buscall1988scaling}. What is clear is that the elasticity in these arrested and disordered amorphous solids emerge from the growth of fractal clusters of particles that eventually percolate into a single network spanning the entire sample \cite{kolb1983scaling, family1986kinetics}. Some of the gel mechanics and their dependence of state variables can be determined through mean field approximations \cite{shih1990second, snabre1996rheology}. However, large variations in mechanical and rheological properties of gels with different attractive interactions can not be described nor explained by local microstructural measures of the system such as coordination number of particles \cite{rocklin2018elasticity, jamali2019multiscale}. Linear elasticity models have been developed based upon contact between clusters that are locally and internally glassy in nature, recovering experimentally measured shear moduli of different gels \cite{whitaker2019colloidal,zaccone2009elasticity, bantawa2021microscopic}. Nonetheless, in both experiments, and in simulations,  it is virtually impossible to identify clusters of particles with confidence, and perform a consequent study to confirm these mean field approximations. One would argue that the key to constructing robust microstructure-property relationships in these systems is identifying particle clusters and their emergent networks, as opposed to a local description of the system at individual particle-level. 

Network science aims at understanding the emergent phenomena often observed in complex systems by focusing on the patterns of connections between the constituent parts of a system instead of focusing on the individual parts \cite{strogatz_explore_complex_networks,laszlo_complex_networks_Statistical_mechanics}. This overall look at the collective behavior has enabled thorough analyses of the structural and dynamical characteristics of complex systems despite the sparsity of data due to spatiotemporal limitations of observing complex systems \cite{newman_structure_function_complex_networks,boccaletti_complex_networks_structure_dynamics}. Thus network science has been providing pivotal tools for understanding the relationship between structure and function of complex systems in various disciplines ranging from biology, medicine, and neuroscience to epidemiology, ecology, and social sciences \cite{network_biology_cell_functional_organization,network_medicine_emerging_paradigm,network_medicine_approach,complex_brain_networks,epidemic_processes_complex_networks,ecological_networks_fragility,computational_social_science}. One of the cornerstones of network science is the classification of groups of nodes with varying size into clustered elements that are similar to each other with respect to common attributes. The calculation of modularity and detection of clustered structures (community detection) can be done in various ways, unveiling hidden characteristics in many social and biological networks \cite{community_structure_social_biological_nets_girvan_newman,newman_finding_evaluating_community_structure_networks,newman_modularity_community_structure_networks}. 

In this work, we leverage advances in network science to accurately identify colloidal clusters within a single giant network of particles, and recover the elastic response of the emergent gels from a coarse-grained cluster network. By doing so, we provide a systematic pathway to recovering mechanics of a complex network, from a single snapshot of a system at quiescent conditions. Our results clearly show a one-to-one correspondence between ``elasticity in particulate systems'' and ``resilience in complex networks''. Furthermore, the analysis of the resulting networks and their corresponding elastic moduli help us identify phase maps from simulations of a single volume fraction of colloidal particles, far beyond the phase boundaries that make experimental and computational studies of the phase diagram challenging.

\section*{Coarse-graining the particulate network into clusters, and elasticity measurements}
\begin{figure}[!h]
    \centering
    \includegraphics[trim={1.5cm 8.5cm 4cm 6cm},clip, width=\textwidth]{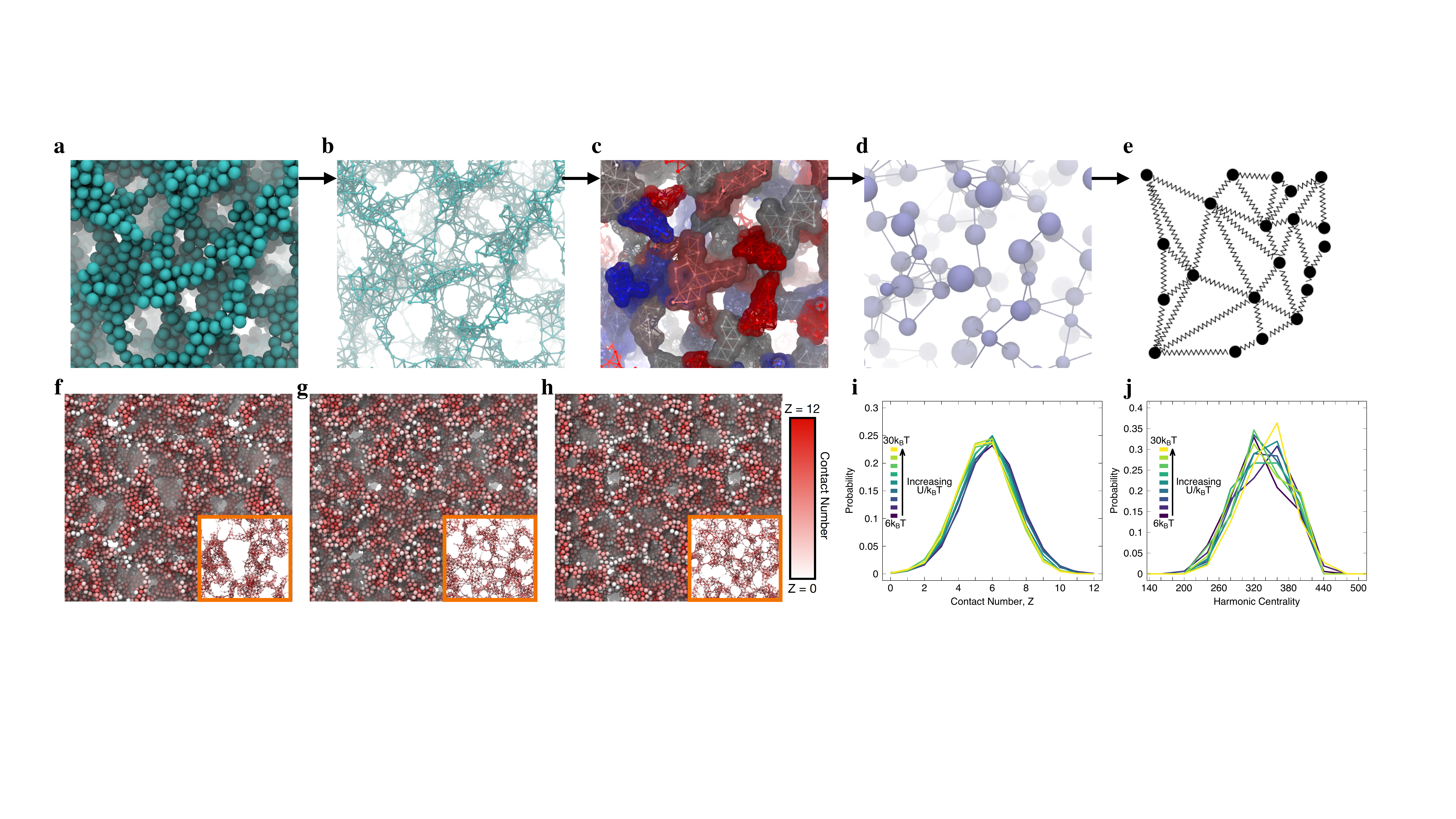}
    \caption{\textbf{Schematic view of the clustering and coarse-graining of the colloidal network.} \textbf{(a)} A magnified snapshot of the colloidal particles after gelation, \textbf{(b)} Network of interparticle bonds, \textbf{(c)} Clustered particles after GMM algorithm with coloring as visual aid,  \textbf{(d)} Coarse grained network, and \textbf{(e)} Spring network model of the coarse grained network.  Snapshots of particles after gelation for \textbf{(f)} $\textit{U}_0 = 6 \textit{k}_{\mathrm{B}} \textit{T}$, \textbf{(g)} $\textit{U}_0 = 12 \textit{k}_{\mathrm{B}} \textit{T}$, and \textbf{(h)} $\textit{U}_0 = 21 \textit{k}_{\mathrm{B}} \textit{T}$; Insets show a portion of the interparticle networks.  \textbf{(i)} Coordination number distribution, and \textbf{(j)}  Harmonic centrality distribution at different attraction levels.}
    \label{Fig1}
\end{figure}

\vspace{-20pt}
\noindent\rule{6.3in}{0.4pt}
The overarching scheme of the network analysis in colloidal gels of interest in this work is shown in Fig. \ref{Fig1}. Accurate large scale Dissipative Particle Dynamics (DPD) simulations are used to model attractive colloidal gels at an intermediate volume fraction. DPD method has been employed previously in order to study the structural features of colloidal gels during the gelation, as well as their rheological characteristics in the linear and non-linear flow regimes \cite{jamali2019multiscale, jamali2017microstructural,jamali2020time,nabizadeh2021life}. The depletion interaction between the colloidal particles leads to formation of thermo-reversible bonds, and eventually into a space-spanning network of particles (Fig. \ref{Fig1}a). We construct colloidal networks from DPD simulations, where nodes represent particles and each particle-particle bond is represented by an edge (Fig. \ref{Fig1}b). Two particles are bonded if their interparticle distance ($r_{ij}$) is small enough for their attraction strength to exceed $5 \textit{k}_{\mathrm{B}}T$.

In the next step, we infer the spatial location of nodes only from the network structure, ignoring the actual spatial coordinates of particles to let the definition of particle-particle bond drive the analysis. To allow for a natural selection of the size, shape, and number of particle clusters as opposed to imposing biased constraints, we identified clusters of particles using  \textit{Gaussian Mixture Model (GMM)}, a spatial clustering method that considers each cluster as a different Gaussian distribution (Fig. \ref{Fig1}c). This is done through an unsupervised exhaustive algorithm, in which the number of clusters in the system as well as each cluster's individual colloids are rigorously identified based on a Bayesian Information Criterion (BIC). By doing so, we ensure that no adjustable parameters are included in the cluster identification (a.k.a. community detection) algorithm used. Next, we shift focus from particles to clusters by building a cluster network, enabling us to characterize the interactions between clusters of particles  (Fig \ref{Fig1}d). Finally, we translate the cluster network into a mass-spring model to calculate the elastic moduli of the colloidal gels (Fig. \ref{Fig1}e). 
Detailed description of all algorithms developed and used throughout this study can be found in the methods section.

\section*{Particle-level analysis}
\vspace{-20pt}
\noindent\rule{6.3in}{0.4pt}
We simulate gels with different attraction levels from $\textit{U}_0 = 6-30 \textit{k}_{\mathrm{B}} \textit{T}$, with a range of $0.1a$ (\textit{a} being the particle radius) at the volume fraction of ($\phi = 20\%$), consistent with reported experiments  \cite{hsiao2012role}. In a series of reports, using the same state variables, roughly an order of magnitude increase is reported for the elastic moduli of the gels as the attraction increases \cite{hsiao2012role, rocklin2021elasticity, whitaker2019colloidal}. Nonetheless, this significant rise of elastic modulus cannot be described through particle-level descriptors of the system such as coordination number and/or fabric tensor \cite{jamali2017microstructural, zhang2019correlated}. Visual inspection of the gels as represented in Fig. \ref{Fig1}f-h for three different strengths of attraction ($6, 12, 21 k_B T$) also does not indicate distinguishable differences in the domain size or porosity of the resulting particulate networks. In Fig. \ref{Fig1}f,g and h particles are color coded based on their coordination number (the number of a particle's contacting neighbors). To further quantify the microstructural features of each gel, the distribution of contacts per particle (i.e. coordination number) at different attraction levels are shown in Fig. \ref{Fig1}i, indicating a rather insignificant difference in the particle-level structure. Note that coordination number of a particle is equivalent to the degree of a node in network analogy. 

Another quality of a network is its level of inter-connectedness and how closely nodes within a network are connected to one another. This feature can be quantified by calculating the harmonic centrality of nodes in the network defined as $hc_i = \sum_{j, j\neq i} 1/d(i,j)$, where parameter $d(i,j)$ is the minimum distance between nodes $i$ and $j$ belonging to the same network \cite{marchiori2000harmony}. This is calculated by finding the shortest path, the minimum number of walks along the network, that has to be taken from node \textit{i} to reach node \textit{j}. Hence, higher harmonic centralities account for higher accessibility of nodes to each other in the network. The distributions of harmonic centrality do not show any systematic differences for gels formed under varying attraction levels (Fig. \ref{Fig1}j). These findings indicate that the characteristics of particle networks alone do not reflect the variations observed in the rigidity of the overall gels. Indeed, this is not surprising since many studies focused on the coordination number of particles and their spatial characteristics also failed to recover the mechanics of colloidal gels. Following the work of Whitaker et al. \cite{whitaker2019colloidal} where the authors found that minimally connected gel clusters correlate with the elasticity of the entire gel, we hypothesize that the appropriate scale describing the mechanics of gels is the mesoscale cluster length scale. As such, we shift our focus from the characteristics of individual particles to identifying and understanding clusters of particles instead.

\section*{Cluster-level analysis}
\vspace{-20pt}
\noindent\rule{6.3in}{0.4pt}

Considering the inadequacy of the particle-level information, as described in the previous section, one will critically need to identify particle clusters in gel networks. Previous work of Whitaker et al.\cite{whitaker2019colloidal} used a \textit{l}-balanced graph theory, to identify clusters of a fixed length scale (from experimental measurement of a correlation length). Nonetheless, experimental reports suggest that reaction-limited aggregation of colloids lead to large polydispersity in cluster size \cite{krall1998internal}. Mass-polydispersity of clusters was also observed via confocal microscopy \cite{dinsmore2002direct}. Topological clustering of computationally simulated gels also result in clusters of varying sizes \cite{royall2015probing}. Theoretically, seminal work of Shih and Shih \cite{shih1990scaling} had established that there exists an average cluster size, that can be used to recover the yield stress or limit of linearity scalings of colloidal gels through a mean field approximation. Indeed, the work of Whitaker and coworkers \cite{whitaker2019colloidal} used this single length scale for clusters in conjunction with a mean-field description (Cauchy-Born theorem) to recover the elasticity of colloidal gels; however, this does not mean that all clusters within the gel structure are mono-sized. These gels are disordered arrested structures with fractal-like topologies that naturally bring about the polydispersity of the clusters.

On the other hand, results in Fig. \ref{Fig1}i suggest that it is safe to assume a Gaussian distribution for the degrees of nodes (coordination numbers) in the network of particles. As these systems represent self-similar structures that hold at smaller scales, it is plausible to assume that each cluster will involve a Gaussian degree distribution within itself. The goal thus is to rigorously and without any adjustable parameters, identify clusters of particles in which a Gaussian distribution is present for the number of particle contacts. In this work, we employ GMM to identify clusters as individual contributions to a total mixture of Gaussian distributions with varying shapes and size. 

In this approach, the optimal number of clusters is identified by minimizing a Bayesian Information Criterion (BIC) function that is recursively calculated for all possible cluster combinations [from one cluster representing the network, down to each cluster having only one particle in its structure]. The BIC values for different cluster numbers are presented in Fig. \ref{SI_BIC}, marking the optimum number of clusters for each attraction strength between the particles. Additionally, the actual spatial configuration of nodes were not considered in identification of clusters. Instead, the spatial configuration of nodes were converted into vectors, allowing for a series of embedded 3-dimensional coordinates using the \textit{Uniform Manifold Approximation and Projection} (UMAP) method. Such dimension reduction and graph learning through Node2Vec are commonly used to learn lower-dimensional embedding of the nodes. A schematic view of the process for cluster identification in this analysis is illustrated in Fig \ref{fig_clustering_pipline}.

We use two different definitions for the cluster diameter to ensure that identified clusters are indeed rigid assemblages of colloids. From the physical stand point, we define physical diameter of a cluster as the diameter of the sphere that contains all individual particles in that cluster, denoted by $D_{Phy.}$. From a network perspective diameter of the cluster can be expressed as the length of the longest shortest path connecting two nodes (the minimum number of walks along edges needed to connect any two nodes within a cluster) in that cluster, $D_{Net.}$.  For the clusters to remain rigid, as the physical diameter grows, so must the network diameter at the same rate. Up to the network distance of eight (8) the physical-distance and network-distance are almost identical (Fig. \ref{fig_SI_physical_network_distance}). This is rather consistent with the correlation length found in \cite{whitaker2019colloidal} for the cluster size and set the largest length scale for the individual clusters identified here. Since clusters with $D_{Net.}>8$ are assumed to no longer be rigid in nature, we apply GMM recursively to partition those clusters into sub-groups of particles with $D_{Net.}\leq 8$. Lastly, upon clustering process, a limited number of clusters of size one and two are also identified (referred to as orphan nodes), which are consequently removed from the remaining analysis. This is to ensure that a collection of cohesively connected particles are included in the study. Note that these nodes account for $\leq0.7\%$ of the particle population for different attraction strengths.

\begin{figure}[h]
    \centering
    \includegraphics[trim={9cm 8cm 9cm 8cm},clip,width=\textwidth]{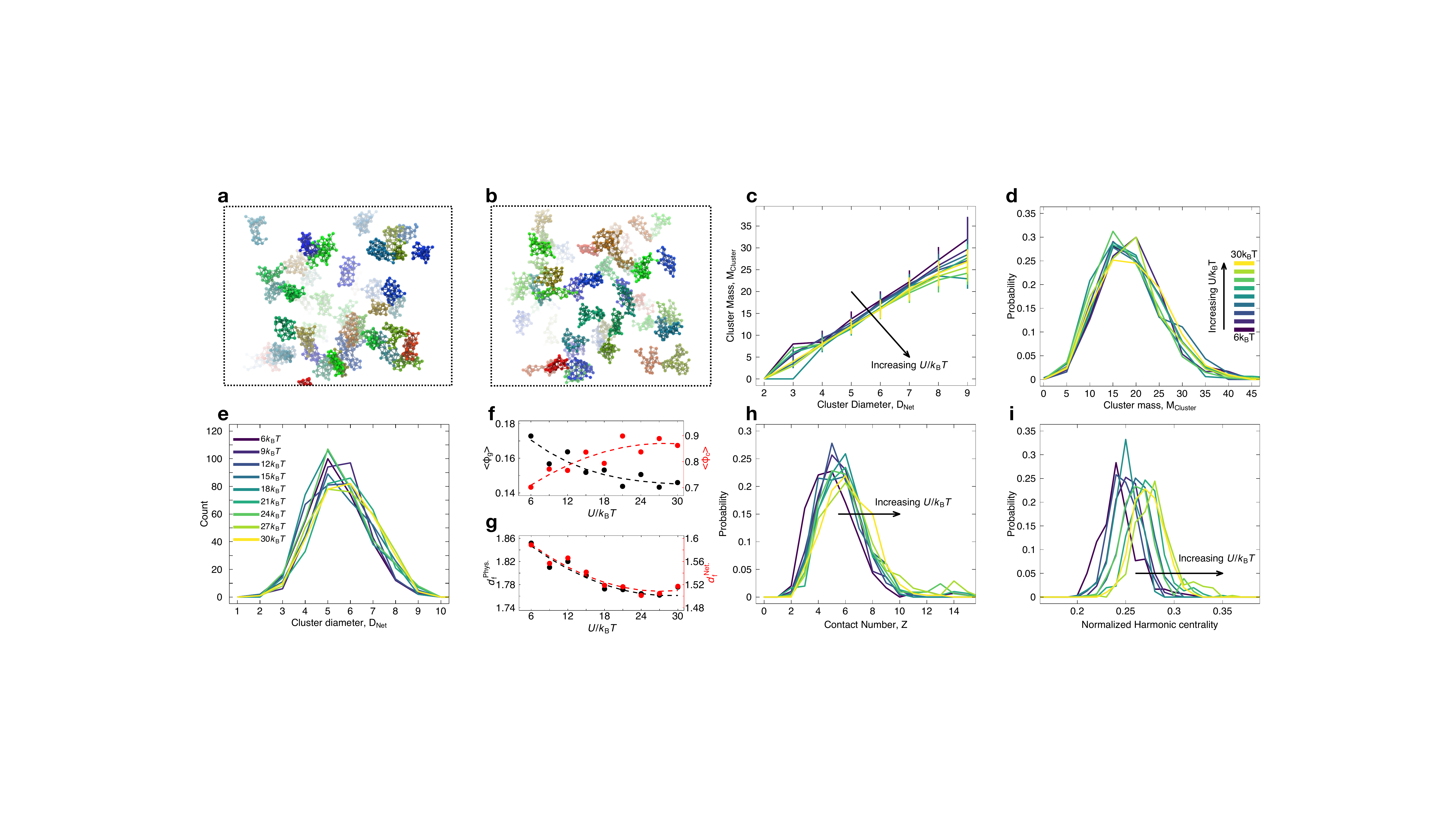}
    \caption{\textbf{Analysis of the coarse grained cluster network.} Snapshots of 40 clusters at \textbf{(a)} $\textit{U}_0 = 6 \textit{k}_{\mathrm{B}} \textit{T}$, and \textbf{(b)} $\textit{U}_0 = 30 \textit{k}_{\mathrm{B}} \textit{T}$. \textbf{(c)} Cluster mass versus cluster diameter.  Distribution of \textbf{(d)} cluster mass, and \textbf{(e)} cluster diameter ($d_{Net.}$) versus attraction level. \textbf{(f)} Internal (black) and external (red) cluster volume fractions versus attraction strength. \textbf{(g)} Fractal dimension of clusters calculated in  physical (black), and network (red) dimensions. 
    \textbf{(h)} Cluster coordination number ($Z$), and 
    \textbf{(i)} Normalized harmonic centrality versus attraction strength.
} 
    \label{Fig2}
\end{figure}

Snapshots of forty randomly selected clusters annotated by our approach for two different attraction strengths are shown in Fig. \ref{Fig2}a and b. Nodes belonging to the same cluster are colored similarly to aid visual inspection of clusters. We observe more compact clusters for the weak gel ($U_0 = 6k_B T$), compared to the strong gel ($U_0 = 30 k_B T$) which shows relatively more elongated clusters in the final gel. This is in agreement with previous descriptions of structural heterogeneity in clusters of more attractive colloids \cite{dibble2006structure}. Considering the monodispersity of colloidal particles in diameter, we define cluster mass as the number of particles in a cluster.  The comparison of the physical cluster mass with the network-based cluster diameter ($D_{Net}$) across all attraction levels also confirms the existence of more elongated clusters in gels formed by higher attraction strengths (Fig. \ref{Fig2}c).  For instance, given a fixed cluster mass, cluster diameter of the gel formed at $U_0=30\textit{k}_B\textit{T}$ is larger than the ones observed in the weaker gels. On the other hand, the distribution of cluster mass (denoted by $M_{Cluster}$) and network-based cluster diameter in Fig. \ref{Fig2}d,e show attraction-independent behavior, indicating that the higher internal cluster interconnectivity in the network of weaker gels does not originate from the mass and diameter differences of clusters.

From a physical perspective, stronger gels yield smaller internal volume fractions, from $\phi_g = 0.18$ for $U_0 = 6 k_B T$ to $\phi_g = 0.14$ for $U_0 = 30 k_B T$. However, clusters of stronger gels are larger in diameter (the smallest sphere that embodies all the particles in the cluster), and the total volume of the clusters grow significantly as the attraction strength increases, reaching fractions of $\phi_{Cluster}>0.9$ (Fig. \ref{Fig2}f). 
Note that since clusters are polydisperse in nature, the fraction of these clusters can easily surpass values measured for glassy regime in monodisperse clusters as proposed by Whitaker et al.\cite{whitaker2019colloidal}.  Clusters represent fractal structures made from individual particles; hence, their internal microstructure can be quantified through a fractal dimension. Here, we define two different measures of the fractal dimension, based on physical measures, and network measures (Fig. \ref{Fig2}g) of the annotated clusters. Mathematically, the physical fractal dimension, $d_f^{Phy.}$, can be defined as $d_f^{Phys.} = \frac{\log (R_{g})}{\log (M_{Cluster})}$, $R_g$ being the radius of gyration of the cluster, and the network fractal dimension, $d_f^{Net.}$, can be written as $d_f^{Net.} = \frac{\log (D_{Net.})}{\log (M_{Cluster})}$.

With the annotated clusters of particles, we construct a cluster network of the gel, in which nodes are clusters connected by an edge if there exist a boundary edge or orphan particle between them. These edges are weighted by the number of boundary edges and orphan particles connecting two clusters given the structure of the particle network. While the particle-level characterization of the gel network does not indicate attraction-dependent properties (Fig. \ref{Fig1}i and j), the cluster network at different attraction levels show distinct features and characteristics. For instance, the degree (i.e., cluster coordination number) and harmonic centrality distributions of clusters in the cluster network show a clear shift towards larger values as the attraction strength increases (Fig. \ref{Fig2}h and i).  That indicates higher inter-connectivity in the cluster network as the attraction strength increases between individual particles. That can be further interrogated through harmonic centrality of the clusters in the cluster network shown in Fig. \ref{Fig2}i. While the harmonic centrality measurement for the network of individual colloids in Fig. \ref{Fig1}j showed no visible differences between different values of attraction strength, the same measure shows a clear and systematic increase of inter-connectivity in the network of clusters.

\section*{Elastic modulus and resilience of a gel network}
\vspace{-20pt}
\noindent\rule{6.3in}{0.4pt}
The mechanics of the constructed cluster networks can be further investigated using a simple spring network model (See SI Methods section). Each cluster (regardless of the diameter) is represented by a mass, and edges are replicated through a spring whose constant reflects the length of the shortest path connecting the cluster pair. In these calculations, we assume that the cluster networks are in mechanical equilibrium, then their elasticity is measured in response to an infinitesimal affine strain deformation using the Born-Huang formulation~\cite{Maloney_PRE_2006}. Note that more complex models, reflecting on the size, shape, and volume fraction of cluster are possible; however, the goal is to assess the ability of a crude cluster network without specific particle-level information to describe the rigidity of colloidal gels.

An important feature of the mass-spring model calculations presented here is the spring constants used to describe the cluster-cluster connections. For approximation of the stiffness of inter-cluster bonds, we use the polymer chain stiffness theorem in which the stiffness of a chain is approximated as a function of both the single particle-particle stiffness, and the size of the backbone of the chain. In particular, stiffness of a cluster-cluster connection of length $d_{ij}$ is estimated as $\textit{K}_s(\textit{U}/ \textit{k}_B \textit{T}, d_{ij}) = -\textit{K}_s(\textit{U}/ \textit{k}_B \textit{T},1) \times  log(\frac{1}{d_{ij}+1}))/log(1/2) $, where $\textit{K}_s(\textit{U}/ \textit{k}_B \textit{T},1)$ is the stiffness of a single particle-level bond and $d_{ij}$ is the lengths of the longest shortest path within the cluster (this is the equivalent of the size of the cluster backbone). The bond stiffness values obtained from this approximation are compared to the experimentally measured results of Dinsmore et al. \cite{dinsmore2002direct} in Fig. \ref{kappa_comp}, showing an excellent tracking of the bond stiffness values from experiments.

\begin{figure}[!h]
    \centering
    \includegraphics[trim={11cm 6cm 8cm 5cm},clip,width=\textwidth]{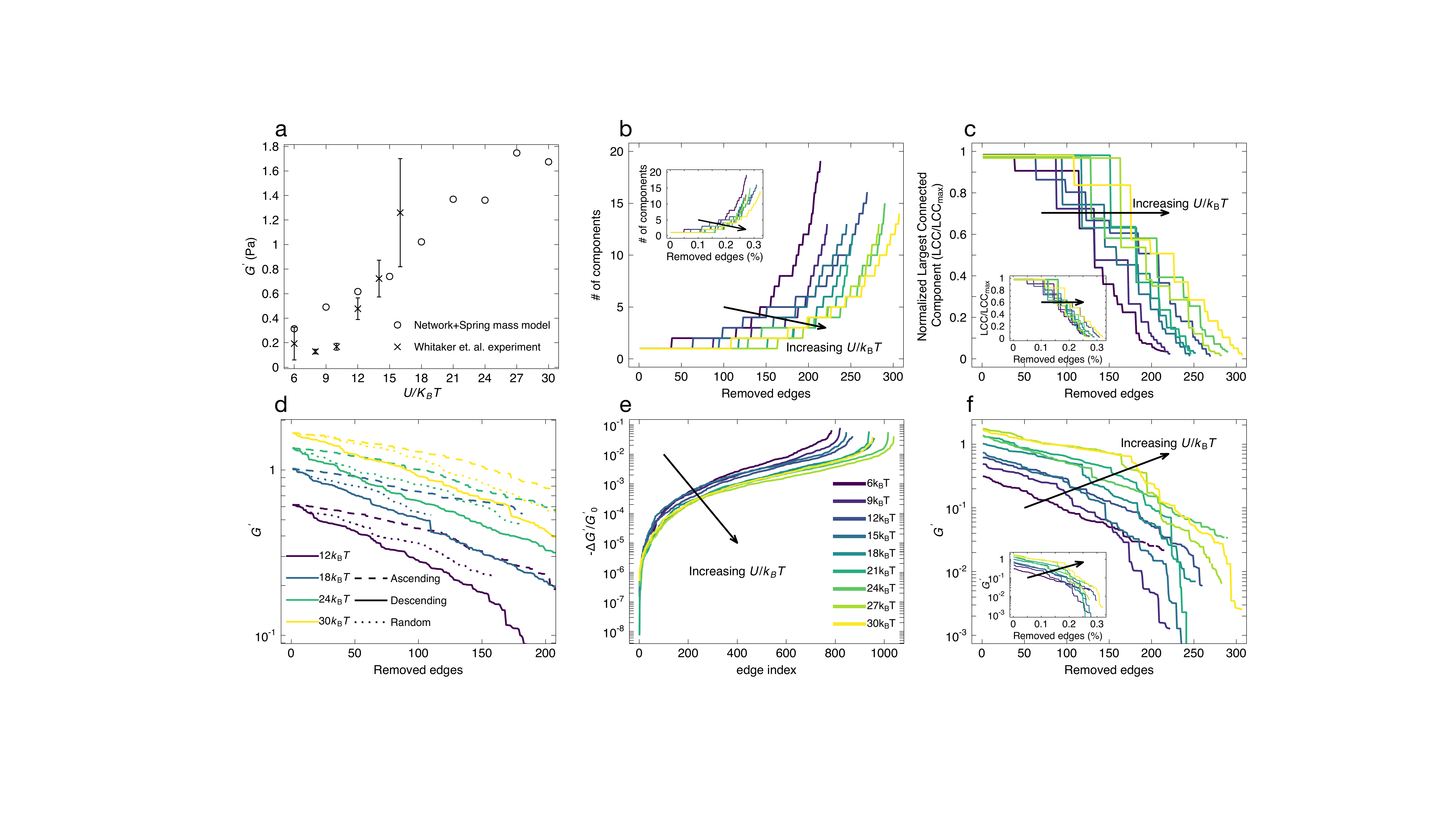}
    \caption{
    \textbf{Elastic modulus and network resilience against edge removal.} 
    \textbf{(a)} Elastic modulus, $G'$, of different colloidal gels versus attraction strength. 
    \textbf{(b)} Number of connected components versus number of removed edges, inset shows number of connected components versus percentage of removed edges.
    \textbf{(c)} The largest connected component (LCC) versus number of removed edges, inset shows LCC versus percentage of removed edges.
    \textbf{(d)} Elastic modulus, $G'$, versus  number of removed edges for different attraction strengths, calculated from three scenarios of edge removal: betweenness-ascending (dashed line), betweenness-descending (solid line) and random (dotted line).
    \textbf{(e)} Elastic modulus loss ($\Delta G'= G'-G'_0$) normalized by the initial value of the elastic modulus before edge removal ($G'_0$) versus bond index. 
    \textbf{(f)} Shear modulus versus the percentage of removed edges in a betweenness-descending manner, inset shows shear modulus versus percentage of removed edges.
}
    \label{Fig3}
\end{figure}

Fig. \ref{Fig3}a shows the elastic shear moduli of the gels at different attraction levels compared to the experimental measurements of the depletion gels at similar system variables (solid fraction, and attraction range/strength)  \cite{hsiao2012role}. Our coarse-grained spring network model recovers the elasticity of the gels quantitatively, strongly suggesting that: (i) our network-based approach identifies particle clusters correctly, and (ii) the cluster-level information is indeed necessary and sufficient for the recovery of rigidity in colloidal gels. Having established that the mesoscale cluster network is reflective of the gel mechanics at the macroscopic level, one can interrogate the cluster network's characteristics and their correlations with the physical properties of colloidal gels. In particular, here we study the resilience of cluster networks and their correlation to the elasticity of colloidal gels.

\section*{Resilience}
\vspace{-20pt}
\noindent\rule{6.3in}{0.4pt}
Used routinely as a key characteristic of many complex systems, resilience is generally defined as a complex system’s ability to retain its basic functionality upon exposure to defaults \cite{error_attack_tolerance_networks,universal_resilience_patterns_networks,resilience_network_motifs,cascade_based_attacks_networks,optimizing_networks_against_cascading_failure}. Defined mathematically based on changes in a particular function over time, as an environmental change is posed, resilience is commonly referred to the point at which non-linear changes in a system's performance is observed. Here, we studied the resilience of colloidal networks as their ability to maintain functional properties upon loss of edges. Hence, the order by which edges are removed from a network can significantly impact its resilience. One approach to appropriately assess resilience in a network is edge removal based on betweenness centrality as it targets the most central edges in providing shortest connections in that network. It should be noted that edge betweenness centrality is primarily introduced in the Girvan-Newman algorithm, a community detection technique for partitioning a network into clusters of cohesively connected nodes \cite{community_structure_social_biological_nets_girvan_newman}. In this algorithm, edges with the highest betweenness centrality are progressively removed until no edges remain (See Methods). The number of connected components in each system against the number of edges removed, for the studied attraction strengths are shown in Fig. \ref{Fig3}b. Note that the algorithm is applied on the cluster network, and thus the total number of nodes identify the number of clusters in the system. For a fixed number of removed edges, cluster networks at higher attraction strengths consistently have a smaller number of connected components, i.e. are more resilient to the removal of central edges. That remains valid even when the number of connected components and removed edges are normalized by the total number of nodes and edges, respectively, in each cluster network (inset of Fig. \ref{Fig3}b). This further suggests that even for the same number of clusters, and the same number of edges between clusters, networks formed at higher attraction strengths are more resilient to loss of cluster-cluster connections. 

While the number of connected components is a measure of how a network reacts to the loss of an edge, it is the largest connected component (LCC) within the system that remains responsible for the elasticity of a gel. Fig. \ref{Fig3}c shows the size of LCCs in cluster networks normalized by the network size, against the number of removed edges. We observe that removal of the first \%5 of connections among the clusters does not change the size of LCCs. Afterwards, LCCs in cluster networks with higher attractions tend to generally be larger than the ones for the lower attractions at any given number of removed edges, i.e. exhibit more resilient behavior. The trends in the inset of \ref{Fig3}c, the size of LCCs against the percentage of total removed edges, further suggests that the resilience of the stronger gels does not solely originate from their higher number of connections.

The dynamic gel property of interest during the resilience study is chosen to be its elasticity. While the higher resilience of gels with higher attraction strength between the particles is qualitatively demonstrated through results in \ref{Fig3}b/c, as mentioned before the order of bond loss (which bond is cut from the network first) is a consequential decision to make. To further test this hypothesis, physical resilience of the cluster networks are studied upon removal of edges in a series of separate simulations. One would expect that loss of different edges will have different effects on the mechanics of the cluster network. To show these significantly different effects on the elasticity of the network, we also performed an exhaustive series of simulations where one single edge is removed from the initial cluster network in each simulation. The loss of elasticity upon each edge removal trial is sorted in an ascending order and presented in Fig. \ref{Fig3}e. These clearly show that loss of elasticity upon elimination of a single connection between clusters can vary over seven orders of magnitude, suggesting that loss of some edges have minimal effect on the networks modulus while other edges' removal can result in detriment of bulk elasticity up to \%8 of its initial value. Further analyzing the edges for which the highest levels of elasticity losses are measured revealed that the betweenness centrality of an edge is significantly correlated with the elasticity loss (Fig. \ref{SI_betc_G_corr}). As the betweenness centrality of an edge reflects its relative role in the transmission of stress across a system, edges of higher betweenness centrality will be more likely to localize stresses and play a crucial role in material failure. In other words, edges with higher betweenness centrality contribute more to the rigidity of a gel network. 

This can be directly examined by a series of resilience studies, in which edges are removed based on different scenarios and the remaining structure's elasticity is measured by the mass-spring model. Fig. \ref{Fig3}d shows the elastic moduli of the networks against the number of removed springs from the system using three edge removal approaches: (1) random, (2) ascending order of betweenness centrality, and (3) descending order of betweenness centrality. These results further confirm that edges with higher values of betweenness centrality are more essential to gels' rigidity as their removal results in a more rapid loss of elasticity. This strongly suggests that the edge betweenness centrality can be used as an indicator of the failure points in the structure of colloidal gels.

Once the appropriate mode of edge removal is established (in descending edge betweenness centrality order), a thorough investigation of the elasticity-resilience correlation can be performed on the cluster networks. In Fig. \ref{Fig3}f, the moduli of the gels were measured for cluster networks as they lost edges within their structure until no elastic response could be recovered for the system. These results once again confirm that gels formed at higher attraction strengths are more resilient to loss of a cluster-cluster spring, as higher elastic moduli are measured for those with the same number of removed edges. This is valid even when normalizing the number of removed springs to the total number of springs in the system (inset to Fig. \ref{Fig3}f).

\section*{Gelation phase boundary.} 
\vspace{-20pt}
\noindent\rule{6.3in}{0.4pt}
In results presented in Fig. \ref{Fig3}, for each edge removal instance, the rigidity is determined by its largest connected component. With the resilience measurement and the elastic moduli calculated from the spring network model, one can find a threshold at which the rigidity emerges in a cluster network. To do so, the actual volume fraction of the largest connected component remaining in the system is measured upon removal of the edges. However, a singular definition to be applied to a gel and identify whether it can be considered ``rigid'' does not exist. Thus, here, and to remain consistent with experimental measurements in \cite{hsiao2012role, whitaker2019colloidal, lu2008gelation}, we chose $\textit{G}^{'} = 0.1Pa$ as the criterion for identifying a network as rigid. 

\begin{figure}[!h]
    \centering
    \includegraphics[trim={4cm 6cm 4cm 3.cm},clip,width=\textwidth]{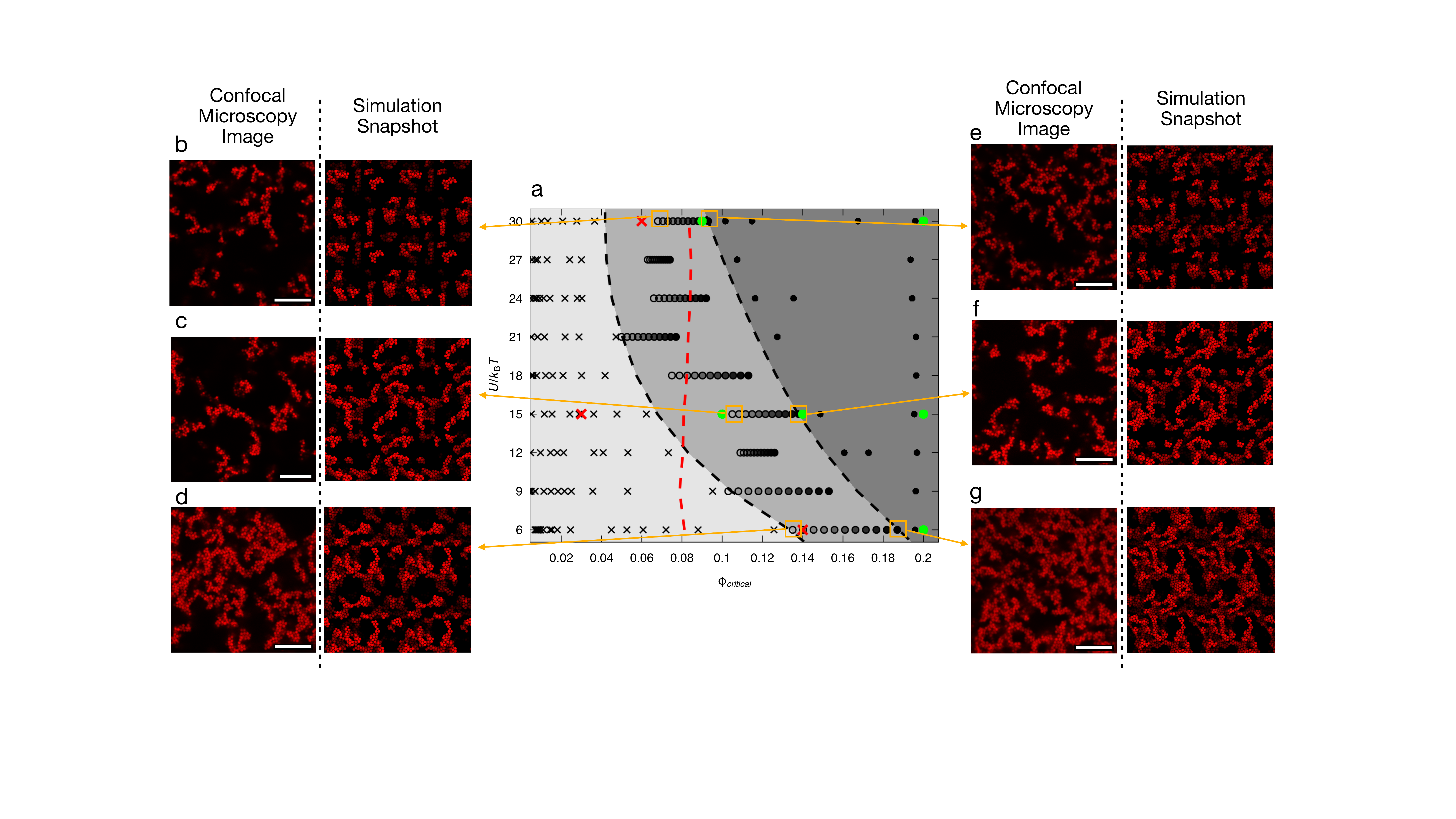}
    \caption{\textbf{Gelation phase boundary.} \textbf{(a)} The minimum and maximum solid fractions required for the emergence of elasticity. Data points are determined from resilience analysis, and the phase boundaries (black dashed lines) show the minimum and maximum volume fractions where  $\textit{G}^{'} \geq 0.1 Pa$. Fluid states are shown by a cross symbol and rigid states are shown by filled circles. The red dashed line represents the percolation line where the average coordination number of a cluster-level network exceeds the critical coordination number $Z_C = 2.4$ \cite{rouwhorst2020nonequilibrium}. Red symbols indicate the experimental results for the fluid (cross) and gel (filled circles) states. Snapshots of the particulate structures are shown for the lower bound of solid fractions that satisfy rigidity at \textbf{(b)} $30 \textit{k}_{\mathrm{B} \textit{T}}$, \textbf{(c)} $15 \textit{k}_{\mathrm{B} \textit{T}}$ and \textbf{(b)} $6 \textit{k}_{\mathrm{B} \textit{T}}$, and also the higher bounds at \textbf{(e)} $30 \textit{k}_{\mathrm{B} \textit{T}}$, \textbf{(f)} $15 \textit{k}_{\mathrm{B} \textit{T}}$ and \textbf{(g)} $6 \textit{k}_{\mathrm{B} \textit{T}}$, compared to experimentally observed structures from the confocal microscopy of PMMA depletion gels at the same system variables. The scale bar in the confocal images is $10 \mu m$.
    } 
    \label{FigF}
\end{figure}

In Fig. \ref{FigF}.a, we plot the gelation phase boundary, measured from the resilience analysis, where the volume fraction, $\phi_{critical}$, is the fraction of colloids in the largest connected component. The lower and upper boundaries of solid fraction are the required volume fractions to satisfy the rigidity condition, $\textit{G}^{'} = 0.1Pa$. This phase diagram is reminiscent of what has been suggested by experiments and theory \cite{lu2008gelation,zaccarelli2007colloidal}, clearly showing a gelation at lower solid fractions for higher attraction strengths. Our results suggest that for strong gels ($U/ k_B T \geq 15$) the minimum volume fraction of $\phi = 0.05$ is required for a rigid gel to emerge, and as soon as a percolated network is formed. On the other hand, for weak gels of ($U/ k_B T < 15$), percolation (the dashed red line) simply does not result in rigidity, and significantly larger fractions of colloids are required for an elastic gel to form. To further validate the predicted gelation phase diagram and the phase boundaries in Fig. \ref{FigF}.a, we experimentally study the gelation behavior of sterically stabilized, charge screened poly(methyl methacrylate) (PMMA) colloids suspended in a solvent containing polystyrene as a short-range depletant. The experimental phase space spans nearly the entire range of U/kT and $\phi$ values shown in Fig. \ref{FigF}. Representative confocal microscopy images of the PMMA colloidal gels at three different attraction strengths and different volume fractions are shown in Fig. \ref{FigF}.b-g, compared to the snapshots from the simulations, showing visually that the network resilience-based reconstruction of the phase diagram is accurate in predicting the gel-fluid states. A side-by-side comparative view of the structure at long times, from the simulations and the confocal imaging is provided in Supplemental Video S1. To further proof the state of particulate structure beyond a visual inspection, microdynamical and microstructural data were obtained from the confocal microscopy images, and are presented in Fig. \ref{Fig5}. We also performed a series of simulations at the same volume fractions as experimentally investigated to ensure that the microstructural and microdynamical evolutions of the system are indeed appropriately captured in our simulation scheme. The results in Fig. \ref{Fig5}.a-d show comparison of van Hove self correlations for three different volume fractions and two different attraction strengths measured experimentally and computationally, showing a close agreement between the two. These self-correlation graphs as well as the experimentally measured ensemble-averaged diffusion of particles show clear differences between ungelled and gelled samples. Specifically, colloidal gels demonstrate kinetic arrest through a significantly reduced mean squared displacement that is independent of lag time, while ungelled particulates and ``clusters of fluids'' states exhibit mostly diffusive motion even at very long times. These are shown clearly in Fig. \ref{Fig5}.e, where the mean squared displacement of particles are plotted against the lag time for a number of different systems. Similarly, the van Hove self-correlations of the particle displacement, obtained from experimental measurements and simulations, both demarcate the gel states from fluid clusters and freely dispersed particles. Videos from the confocal microscopy showing the different structures of the colloidal systems in gelled, clusters of fluid, and diffusive states are provided in Supplemental Videos S2-4 respectively, clearly indicating that the structures observed are indeed stable at long times.

\section*{Conclusions}
\vspace{-20pt}
\noindent\rule{6.3in}{0.4pt}
We have shown through a series of detailed particle-level simulations, network analyses and spring-network modeling benchmarked and validated against experimental measurements that the general mechanics of colloidal gels as space-spanning networks of attractive colloids can be studied with respect to their network characteristics. We adapted a Gaussian Mixture Model (GMM) methodology to annotate rigid clusters formed at the mesoscale, and showed that cluster-level networks exhibit distinct features not detectable at particle-level networks. Namely, particle clusters show an increased number of cluster-cluster connections and harmonic centralities as the strength of attraction between individual particles increases. These polydispersed fractal clusters can occupy up to $90\%$ of the entire sample volume, with decreasing internal volume fraction at higher strengths of attraction. The physical- and network-based fractal dimensions of annotated clusters are also consistent with the theoretical mean field predictions. We then showed that a simple mass-spring model of the cluster networks can recover elastic moduli of the gels quantitatively, compared with the experimental measurements.

To measure elasticity of a system, one needs to study dynamic response of that system to an applied deformation. We showed that elasticity of a gel network is correlated with its resilience. This is significant, as resilience of a network can be interrogated from snapshots of a system, without a need for dynamical information. Hence, one can use resilience of a gel network as a proxy to its elasticity. More importantly, these resilience analyses enabled us to construct a fully resolved phase diagram for colloidal gelation, from a series of simulations at a single colloidal volume fraction well beyond the solid-liquid phase boundary. Further validation of the network-predicted phase diagram through experiments shows that the phase boundary and the gel/fluid states recovered from the network and resilience analysis are indeed observed experimentally. This is further demonstrated using detailed calculations of the mean squared displacement of particles at different volume fractions and attraction strengths, as well as van Hove self correlations of the examined attractive colloidal systems. In practice, this means that with a very few selected experiments/simulations resolved at the particle level, and employing these network investigations one can construct detailed state diagrams without exploring the entire phase space. Even though our results are based on short-range attractive colloids, we believe our methodology is applicable to a wide range of particulate systems well beyond colloidal gels.

\section*{Methods}
\vspace{-20pt}
\noindent\rule{6.3in}{0.4pt}
\subsection*{Dissipative Particle Dynamics, DPD, simulations} Dissipative Particle Dynamics (DPD) is a discrete model, formulated to simulate the motion of a fluid through explicit pairwise interactions. 

The equation of motion for the DPD method is as follows:
\begin{equation} \label{eq:DPD_equation_motion}
\textit{m}_i \: \frac{\text{d}\textbf{v}_\textit{i}}{\text{d} \textit{t}} = \sum_{\textit{i},\:\textit{i} \neq \textit{j}}^{\textit{N}_\textit{p}} \left( \textbf{F}_{\textit{ij}}^C + \textbf{F}_{\textit{ij}}^D + \textbf{F}_{\textit{ij}}^R + \textbf{F}_{\textit{ij}}^H + \textbf{F}_{\textit{ij}}^M\right)
\end{equation}

The background solvent particles interact through the first three terms on the right hand side of eq.\ref{eq:DPD_equation_motion}, where ${\textbf{F}_{\textit{ij}}^C}$ , ${\textbf{F}_{\textit{ij}}^D}$ , ${\textbf{F}_{\textit{ij}}^R}$ represent the pairwise conservative, dissipative and random forces respectively and are calculated as follows.

\begin{equation} \label{eq:DPD_random}
\textbf{F}_{\textit{ij}}^R = \sigma_{\textit{ij}} \omega_{\textit{ij}}(\textit{ij}_{\textit{ij}}) \Theta_{\textit{ij}} \Delta \textit{t}^{-1/2} \textbf{e}_{\textit{ij}}
\end{equation}
\begin{equation} \label{eq:DPD_dissipative}
\textbf{F}_{\textit{ij}}^D = \gamma_{\textit{ij}} \omega_{ij}^2(\textit{r}_{\textit{ij}}) (\textbf{v}_{\textit{ij}}.\textbf{e}_{\textit{ij}}) \textbf{e}_{\textit{ij}}
\end{equation}
\begin{equation} \label{eq:DPD_conservative}
\textbf{F}_{\textit{ij}}^C = \textit{a}_{\textit{ij}} \omega_{\textit{ij}}(r_{\textit{ij}}) \textbf{e}_{\textit{ij}}
\end{equation}
\begin{equation} \label{eq:DPD_weight_function}
\omega_{\textit{ij}} = (1-\textit{r}_{\textit{ij}}/\textit{r}_c)
\end{equation}

The canonical ensemble is formed through the random and dissipative forces where the fluctuation-dissipation requirements is satisfied in connection with those. The random force eq.\ref{eq:DPD_random} introduces thermal fluctuations via a random function, $\Theta_{\textit{ij}}$. Those fluctuations are then dissipated by the dissipative force eq.\ref{eq:DPD_dissipative} that acts against the relative motion of particles $\textbf{v}_{\textit{ij}}=\textbf{v}_\textit{ij} - \textbf{v}_\textit{ij}$. The strength of dissipaton is determined via $\gamma_{\textit{ij}}$ which is coupled with the thermal noise, $\sigma_{\textit{ij}}$. The dimensionless temperature is then determined from the random and dissipative terms $k_\textit{B}T=\sigma_{\textit{ij}}^2/2\gamma_{\textit{ij}}$. $\Delta \textit{t}$ is the simulation time step and $\textbf{e}_{\textit{ij}}$ is the unit vector for interparticle distance. Finally, the chemical identity of a particle based on its chemical potential/solubility in the system is determined through conservative force eq.\ref{eq:DPD_conservative}, where $\textit{a}_{\textit{ij}}$ is the conservative parameter. The random, dissipative and conservative forces are explicit functions of interparticle distance through a weight function (eq.\ref{eq:DPD_weight_function}).
 
The solvent particles and colloidal particles also interact through the same three forces. Furthermore, for the colloid-colloid interactions, the conservative forces are excluded and two other terms are introduced instead. First, a hydrodynamic force $\textbf{F}^H$ is solved for the particles of the solid phase and is formulated as follows:
\begin{equation}
    \textbf{F}^H_{\textit{ij}} = \mu_{\textit{ij}}^H (\textbf{v}_{\textit{ij}} \cdot \textbf{e}_{\textit{ij}}) \textbf{e}_{\textit{ij}},
\end{equation} 
 $\textbf{F}^H$ represents a short-ranged lubrication force and depends on the drag term where $\textit{h}_{\textit{ij}}$ represents the surface-surface distance between two colloidal particles. $\mu_{\textit{ij}}=3 \pi \eta_0 \textit{a}_i \textit{a}_j / 2 \textit{a}_{\textit{ij}}$ is the pair drag ter where $\textit{a}_1$ and $\textit{a}_2$ are the radii of the interacting colloids. In addition to the hydrodynamic force, the interparticle attraction between colloidal particles is modeled via A short-ranged attractive potential \cite{jamali2015microstructure}. Specifically, Morse potential is used to induce attraction and is calculated as follows:
\begin{equation}
    U_{Morse} = \textit{U}_0 (2e^{-\kappa \textit{h}_{\textit{ij}}} - e^{-2\kappa \textit{h}_{\textit{ij}}}),
\end{equation}
where $\textit{U}_0$ determines the depth of attraction well and $\kappa^{-1}$ is the range of attraction. 
\vspace{0.25cm}

\vspace{0.25cm}

\subsection*{Node representation in latent space}  To represent colloidal particles in a lower dimensional space, 3D in our study, we initially applied the node2vec model to obtain a representation matrix for particles and then reduced their dimensions to three by a non-parametric manifold learning technique called UMAP. Both models are addressed in details in the following subsections. 

\textbf{node2vec.} node2vec is a semi-supervised algorithm that utilizes a random walk-based and stochastic gradient descent approaches to learn feature representation of nodes in a network. To do so, it defines a network neighborhood set for every node in the network through a fixed-length second order random walk sampling strategy guided by two parameters $p$ and $q$. These parameters control how fast a walk explores the neighborhood of the starting node (In this study, $p$ and $q$ are set to their default values, i.e., one). Assume we attempt to define a network neighborhood for node $t$ in an unweighted graph. If node $v$ is visited in an initial random walk from $t$, transition probability from $v$ is set by the following rule which incorporates the distance of $t$ to the neighbors of $v$. Traversed nodes after limited number of iterations (controlled by walks per node parameter) are labeled as network neighborhoods of $t$. 

\begin{equation*}
    \alpha_{pq}(t,x) = 
    \begin{cases}
        \frac{1}{p} & \text{if $d_{tx}=0$} \\
        1 & \text{if $d_{tx}=1$} \\
        \frac{1}{q} & \text{if $d_{tx}=2$}
    \end{cases}
\end{equation*}

After choosing a network neighborhood set for every nodes in the network, node2vec tries to maximize the log probability of observing a neighbor for a node conditioned on its feature vector. To do so, it implements stochastic gradient descent in the following objective function, where $f$ are feature representation vectors, $V$ is the set of nodes in the network, $N_S(u)$ is a network neighbor set for node $u$, and $Z_u=\sum_{v\in V}exp(f(u) \cdot f(v))$. 

\begin{equation*}
    \max_f \quad \sum_{u\in V} \big( -log Z_u +\sum_{v\in N_S(u)} f(v) \cdot f(u) \big)
\end{equation*}
\vspace{0.25cm}

\subsection*{Uniform Manifold Approximation and Projection (UMAP)} UMAP is a new manifold learning technique for dimensionality reduction that works in two steps. In the first step, it constructs a fuzzy simplicial complex with the Riemannian geometry theoretical framework. The outcome is a weighted graph describing the manifold structure of data which is then passed to a forced-directed graph layout algorithm to generate a layout in a lower-dimensional space. It starts by defining an open set for every data point and assigning a weighted edge between two overlapping open sets. Open sets are $n$-dimensional spheres with a radius of one concerning a local distance function tuned to included $k$ nearest neighbors of a point. As a result, the edge weight between two data points $x_i$ and $x_j$ is $a+b-a\times b$, where $a$ is a geodesic distance of $x_i$ on the Riemannian manifold of $x_j$, and $b$ is a geodesic distance of $x_j$ on the Riemannian manifold of $x_i$. 

In the next step, UMAP applies a set of attractive (function \ref{attractive}) and repulsive forces (function \ref{repulsive}) to a sample of nodes and edges iteratively to optimize the edgewise cross-entropy between the weighted graph in the first step and an equivalent weighted graph constructed from points embedded in the dimension of interest (denoted by $Y$). $Y$ is initialized by the eigenvector of the normalized Laplacian matrix of the fuzzy graph constructed in the first step.

\begin{equation}
    \frac{
    -2ab\lVert y_i-y_j\rVert_2^{2(b-1)}
    }{
    1+\lVert y_i-y_j\rVert_2^2
    }w_{(x_i,x_j)}(y_i-y_j)
    \label{attractive}
\end{equation}

\begin{equation}
    \frac{2b}{(\epsilon + \lVert y_i-y_j\rVert_2^2)(1+a\lVert y_i-y_j\Vert_2^{2b})}(1-w_{(x_i,x_j)})(y_i-y_j)
    \label{repulsive}
\end{equation}
\vspace{0.25cm}

\subsection*{Gaussian Mixture Models (GMM)} Gaussian mixture model or GMM is the most widely used mixture model that assumes each base distribution is a multivariate Gaussian with unknown parameters (mean and covariance). In case of having $k$ different distributions with mixing coefficient of $\pi_i$ (parameters $\pi_i$, $\forall i\in\{1,...,k\}$ indicate contribution of every model to the overall distribution satisfying $0 \leq \pi_i \le 1$ and $\sum_{i=1}^{k} \pi_i = 1$), the marginal distribution of point $x_n$ is a Gaussian distribution of the following form.

\begin{equation}
    p(x_n) = \sum_{i=1}^{k}\pi_i\mathcal{N}(x_n|\mu_i, \textstyle\sum_{i})\label{gmm}
\end{equation}

With equation (\ref{gmm}), conditional probability (can also be seen as responsibility) of cluster $i$ for explaining data point $x_n$ is computed by the following equation. 

\begin{equation}
    \gamma(z_{ni}) = \frac{p(z_i=1)p(x_n|z_i=1)}{\sum_{j=1}^{k}p(z_j=1)p(x_n|z_j=1)} = \frac{\pi_i\mathcal{N}(x_n|\mu_i, \textstyle\sum_{i})}{\sum_{j=1}^{k}\pi_j\mathcal{N}(x_n|\mu_j, \textstyle\sum_{j})} \label{Bayes'}
\end{equation}

Given equation (\ref{Bayes'}), we use expectation-maximization algorithm to fit a mixture of $k$ Gaussians to representations of $N$ colloidal nodes in a lower dimensional space. This can be achieved by maximizing $L(\mu, \textstyle{\sum}, \pi) = \sum_{n=1}^{N}ln\big(\sum_{i=1}^{k}\pi_i\mathcal{N}(x_n|\mu_i,\textstyle{\sum_{i}})\big)$ taking the following steps. 
\begin{enumerate}
    \item [i]: choose initial values for $\mu_i,~\textstyle{\sum_{i}},~\pi_i~\forall i\in\{1,...,k\}$, and evaluate the initial value of the objective function $L(\mu, \textstyle{\sum}, \pi)$ 
    \item [ii]: re-estimate parameters with the following equations obtained from setting derivatives of the objective function to zero. Note that $m_i=\sum_{n=1}^{N}\gamma(z_{ni})$ estimating the number of points assigned to cluster $i$.
    \begin{itemize}
        \item [-] $\mu_i = \frac{1}{m_i}\sum_{n=1}^{N}\gamma(z_{ni})$
        \item [-] $\textstyle{\sum_{i}} = \frac{1}{m_i}\sum_{n=1}^{N}\gamma(z_{ni})(x_n-\mu_i)(x_n-\mu_i)^{T}$
        \item [-] $\pi_i = \frac{m_i}{N}$
    \end{itemize}
    \item [iii]: re-evaluate the objective function. If the convergence criterion is not met, return to step (ii)
\end{enumerate}

To choose the optimal number of clusters, we run GMM across a wide range of values for $k$ and select the one that minimizes the Bayesian Information Criterion (BIC) function given in equation (\ref{BIC}). $L^{\star}(\mu, \textstyle{\sum}, \pi)$ is the maximum log-likelihood of the estimated Gaussian mixture model for the corresponding $k$.

\begin{equation}
    BIC(k, \mu, \textstyle{\sum}, \pi) = kln(N) - 2\times L^{\star}(\mu, \textstyle{\sum}, \pi) \label{BIC}
\end{equation}
\vspace{0.25cm}

\subsection*{Girvan-Newman algorithm} Edge betweenness centrality of an edge is the sum of the fraction of all shortest paths between two nodes in the network passing through that edge as addressed in equation (\ref{EBC}). 
This centrality was initially proposed in \cite{community_structure_social_biological_nets_girvan_newman} to identify cohesive communities by dropping most central edges in the network. Their method is known as Girvan-Newman algorithm and consists of the following four steps: 
\begin{itemize}
    \setlength\itemsep{-0.5em}
    \item [(i)] Compute edge betweenness centrality for all edges in the network
    \item [(ii)] Remove an edge with the highest betweenness centrality 
    \item [(iii)] Recompute edge betweenness centrality for remaining edges in the network
    \item [(iv)] Repeat steps (ii) and (iii) until no edges remain
\end{itemize}

\begin{equation}
    BC(e) = \sum_{s,t \in V}{ \frac{\sigma(s, t|e)}{\sigma(s, t)} }\label{EBC}
\end{equation}

\subsection*{Calculating the mechanical response of a 3D Mass-Spring network} We consider the system as a 3D spring network under periodic boundary conditions and introduce the potential energy functional for our networks~\cite{Broedersz_Nat_Phys_2011,Rens_EPJE_2019}.
\begin{equation}
E = \sum_{<i,j>} \frac{K_s^{ij}}{2}(L_{ij}-L_{ij}^0)^2 + \sum_{<i,j,k>} \frac{K_b^{ijk}}{2}(\theta_{ijk}-\theta_{ijk}^0)^2,
\end{equation}
where $K_s^{ij}$ and $K_b^{ijk}$ denote the bond-stretching and bond-bending stiffnesses, respectively. $L_{ij}$ represents the edge shared by node $i$ and $j$. $\theta_{ijk}$ is the angle formed by the edge pair $L_{ij}$ and $L_{jk}$. $L_{ij}^0$ is the rest length and $\theta_{ijk}^0$ is the rest angle obtained from the initial configuration, therefore the springs are all in their rest length and the systems are in mechanical equilibrium.

At the system level, we characterize its mechanical response by computing the linear response elastic modulus $G$ to an infinitesimal affine strain $\gamma$ via the Born-Huang approximation~\cite{Maloney_PRE_2006}
\begin{equation}
  G = G_{\text{affine}}-G_{\text{non-affine}} = \frac{1}{V} \left[\frac{\partial^{2}E}{\partial \gamma^{2}}  - \Xi_{i\mu}M^{-1}_{i\mu j\nu}\Xi_{j \nu} \right]_{\gamma=0}.
  \label{eq:sm}
\end{equation}
In Eq.~\eqref{eq:sm}, 
$\Xi_{i\mu}$ is the derivative of the  force on node $i$ with respect to strain given by 
\begin{equation}
\Xi_{i\mu} \equiv \frac{\partial^{2}E}{\partial \gamma \partial r_{i\mu}},
\end{equation}
where $r_{i\mu}$ is the position of node $i$ and $\mu=x,y$ is the Cartesian index. 
$V$ is the total volume of the system. 
$M$ is the Hessian matrix given by the second derivative of the energy $E$ with respect to position vectors of nodes $i$ and $j$
\begin{equation}
M_{i\mu j\nu}=\frac{\partial^{2}E}{\partial r_{i\mu} \partial r_{j\nu}}.
\end{equation}
In our calculations, the bending elasticity $K_b$ is not independent of $K_s$. It's taken as the algebraic mean of $K_s$ of the two neighboring edges $K_b^{ijk} = \kappa \sqrt{K_s^{ij} K_s^{jk}}$, where $\kappa$ is the ratio to manipulate the relative strength between bond-stretching and bending elasticity. The stretching elasticity $K_s$ is calculated based on the number of particle-level connections between the cluster pair.  Inspired by the stress transmission coefficient of a polymeric chain, stretching coeffcient of a cluster-cluster connection is calculated as $\textit{K}_s(\textit{U}/ \textit{k}_B \textit{T}, d_{ij}) = -\textit{K}_s(\textit{U}/ \textit{k}_B \textit{T},1) \times  log(\frac{1}{d_{ij}+1}))/log(1/2) $, where $\textit{K}_s(\textit{U}/ \textit{k}_B \textit{T},1)$ and $d_{ij}$ are the bond-stretching stiffness of a single particle-level bond and the length of connection, respectively.

\subsection*{Synthesis of PMMA colloidal gels}
All chemicals were purchased from Sigma-Aldrich unless otherwise specified. The particles used in this experimental study were poly12-hydroxystearic acid (PHSA) stabilized polymethylmethacrylate (PMMA) colloids prepared using free radical polymerization based on the procedure described by Pradeep et al. \cite{pradeep_contact_criterion}. The particles were dyed with fluorescent Nile Red (peak emission wavelength $\lambda_{em}$ = 635 nm, peak excitation wavelength $\lambda_{ex}$ = 559 nm) for confocal microscopy imaging. The particles were cleaned with pure hexane six times by centrifugation at 10,000 rpm for 15 minutes and stored as dry particles until further use. The particle diameter is 2a = 837 nm $\pm$ 5\% based on the images collected using scanning electron microscopy. The particles were dispersed in a 66:34 volume \% mixture of cyclohexyl bromide (CHB) and decalin containing 1 $\mu\mathrm{M}$ tetrabutyl ammonium chloride (TBAC) to ensure charge screening as well as density and refractive index-matching. To introduce attractive interactions between colloids, we suspended polystyrene (molecular weight $M_{w}$ = 900,000 g/mol, overlap concentration $c^{*}$ = 10.8 mg/mL, radius of gyration $R_{g}$ = 32 $\pm$  2 nm) in the CHB/decalin mixture as a non-adsorbing depletant \cite{whitaker2019colloidal}. Using this method, we prepared colloidal gels with a range of volume fractions ($0.03 \le \phi \le 0.20$) and depletant concentrations ($\textit{c}/\textit{c}^{*}$ = 0.79, 1.75, and 3.35). To estimate the pairwise net potential $\textit{U}$ between the colloids in the gel network, we summed the attractive contribution, computed using the Asakura-Oosawa relation \cite{asakura_oosawa_potential}, and the repulsive contribution computed using the Yukawa potential \cite{yukawa_potential}. The colloidal gel interactions corresponded to $\textit{U}$ = 6, 15, and 30 $\textit{k}_{\mathrm{B}} \textit{T}$.

\subsection*{Confocal imaging and image processing}
Colloidal gels were imaged using an inverted confocal laser scanning microscope (Leica TCS SP8) equipped with a 63$\times$ oil immersion objective. The excitation wavelength of the laser was set to 552 nm. The freshly prepared colloidal gels were placed into a glass vial and loaded onto the microscope (waiting time $t$ = 0). To match the diffusion time steps used in the simulations ($t = 500 \tau _{D}$), we collected 2D time-series images of the gels at $t$ = 6 mins using a resonant scanner (lag time $\Delta t$ = 0.047 s for a total duration of 18.6 s). The total image resolution was 512$\times$512 with a pixel size of 50.01$\times$50.01 $\mathrm{nm}^{2}$. To avoid wall effects, we imaged the gels at a minimum of 15$\mu\mathrm{m}$ above the coverslip. 
Microdynamics of the colloidal gels were analyzed using a brightness-weighted centroid detection and trajectory linking algorithm \cite{crocker1996methods,furst2017microrheology}. The method involves the identification of particle centers based on the brightest pixel followed by subpixel refinement based on the maximum of the local intensity spectra, and linking of particles between each frame in the time series. The particle trajectories were then used to obtain the mean squared displacement (MSD) and histogram of displacements as a function of $\Delta\textit{t}$. In order to limit statistical error in the dynamical parameters to less than $3\%$, the MSD analysis was limited to lag times for which the number of observations is $O(10^{3})$.

\section*{Data Availability}
\vspace{-20pt}
\noindent\rule{6.3in}{0.4pt}
Source Data are provided with this paper and additional data that support the findings of this study are available from the corresponding authors upon reasonable request.

\section*{Code Availability}
\vspace{-20pt}
\noindent\rule{6.3in}{0.4pt}
Simulations are performed using HOOMD-blue, the open source molecular dynamics simulation toolkit, which is publicly available at the developers' website:\\ \href{http://glotzerlab.engin.umich.edu/hoomd-blue/}{http://glotzerlab.engin.umich.edu/hoomd-blue/}

\section*{Acknowledgements}
\vspace{-20pt}
\noindent\rule{6.3in}{0.4pt}
M.N. and S.J. acknowledge support by NSF DMR-2104869. X.L. and D.B. acknowledge support by  NSF  DMR-2046683, MathWorks, the Center for Theoretical Biological Physics NSF PHY-2019745, and Northeastern University's TIER 1 Grant. R.W., Y.S. and L.H. acknowledge support by NSF DMR-2104726 and NSF CBET-1804462. This work was supported and performed on resources made available through Northeastern University's Discovery Cluster.

\section*{Author Contributions}
\vspace{-20pt}
\noindent\rule{6.3in}{0.4pt}
M.N., B.R., and S.J. designed the research; M.N. collected the simulation data; R.W., Y.S. and L.H. designed,  performed, and analyzed experiments; M.N., F.N. X.L., D.B, B.R., and S.J. analyzed data; M.N., F.N., and B.R. performed network analysis; M.N., X.L., D.B., and S.J. contributed to theory; and all authors wrote the manuscript.

\section*{Competing Interests}
\vspace{-20pt}
\noindent\rule{6.3in}{0.4pt}
The authors declare no competing interests.

\bibliographystyle{unsrt}
\bibliography{main.bib}

\blankpage

\setcounter{page}{1}
\resetlinenumber

{\Large{SUPPLEMENTARY INFORMATION}}
\noindent\rule{6.3in}{0.4pt}
\beginsupplement

\begin{center}
\Large{Network physics of attractive colloidal gels: Resilience, Rigidity, and Phase Diagram}
\end{center}

\begin{center}    
    Mohammad Nabizadeh$^{1,*}$, \and Farzaneh Nasirian$^{2,3}$\footnote{Authors equally contributed.}, \and Xinzhi Li$^{3}$, \and Yug Saraswat$^{4}$, \and Rony Waheibi$^{4}$, \and Lilian C. Hsiao$^{4}$, \and Dapeng Bi$^{3}$, \and Babak Ravandi$^{2,3,\dagger}$, \and Safa Jamali$^{1,}$\footnote{Corresponding authors e-mail: bk.ravandi@gmail.com and s.jamali@northeastern.edu}

    date{
    $^1${\small Department of Mechanical and Industrial Engineering, Northeastern University, Boston, MA, 02215, USA} 
    \vspace{-7pt}\\
    $^2${\small Network Science Institute, Northeastern University, Boston, MA, 02215, USA}
    \vspace{-7pt}\\
    $^3${\small Department of Physics, Northeastern University, Boston, MA, 02215, USA}
    $^{4}${\small Department of Chemical and Biomolecular Engineering, North Carolina State University, Raleigh, NC, 27606, USA}
        \vspace{-7pt}
     \\[2ex]%
    }
\end{center}

\vspace{20px}
\section{Simulation Parameters}
\vspace{-20pt}
\noindent\rule{6.3in}{0.4pt}
Dissipative particle dynamic simulations are performed for $~6000$ colloidal particles to recover $\phi = 0.2$ of solid particles in a cubic box, where simulation box is size $50$ times the particle radius ($\textit{a} = 1$) in all directions. Furthermore, $~300,000$ solvent particles are introduced inside the box to reproduce a number density of ($\rho=3$) at dimensionless temperature of ($k_{\mathrm{B}} \textit{T} = 0.1$). Density of colloidal particles is then matched by setting the mass of a colloidal particle to ($\textit{m}_C = 4/3 \rho \pi \textit{a}^3$), with a unit mass for solvent particles ($\textit{m}_S = 1.0$). Overall, $9$ attraction strengths are simulated where the attraction potential is set to $\textit{U}_0 = 6 \textit{k}_{\mathrm{B}} \textit{T}$ for the weakest gel, and $30\textit{k}_{\mathrm{B}} \textit{T}$ for the strongest gel with increasing steps of $3\textit{k}_{\mathrm{B}} \textit{T}$. $\kappa = 30$ determines the range of attraction and replicates a range of $0.1 \textit{a}$ to achieve a short-ranged attraction. Simulations are performed for a Newtonian medium with viscosity $\eta_0$ for $500$ diffusion time steps $\tau_D = 6 \pi \eta_0 a^3 / \textit{k}_{\mathrm{B}} \textit{T}$ to achieve a space spanning network at quasi-steady state.

\section{Unsupervised Network Clustering Pipeline} 
\vspace{-20pt}
\noindent\rule{6.3in}{0.4pt}

\begin{figure}[!h]
    \centering
    \includegraphics[width=\columnwidth]{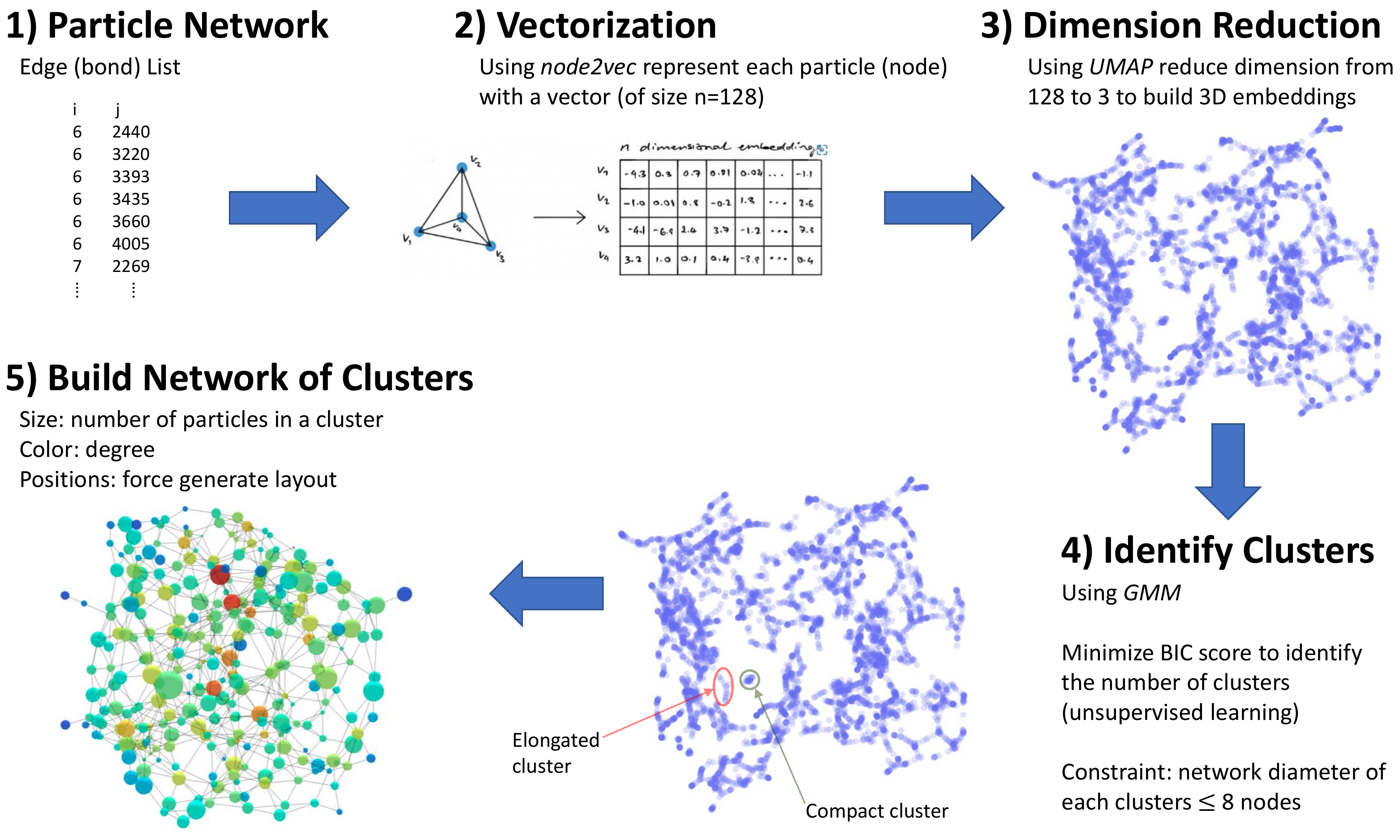}
    \caption{Overarching scheme for the unsupervised network clustering methodology.}
    \label{fig_clustering_pipline}
    \vspace{10px}
\end{figure}

The overall flow chart for the transformation of particulate network into a series of clusters are shown in Figure \ref{fig_clustering_pipline}. In the first step, the list of particle identiciation numbers and their corresponding neighbors are given to a node2vec algorithm for vectorization. This is followed by dimension reduction and community detection. Here GMM was chosen based on the self-similar nature of the gel structure, with a clear Gaussian-like degree distribution for the entire gel. More importantly, there is a rigorous way of finding the subsequent parameters associated with GMM, and no assumption was made here in finding those parameters. These were found instead, through iterative and exhaustive computation. We describe in details all the steps taken in performing these calculations. Figure \ref{fig_clustering_pipline} illustrates the clustering pipeline that we tailored to identify meaningful heterogeneous clusters in colloidal networks. The proposed clustering pipeline is unsupervised since the number of clusters, the most important parameter, is systematically derived based on the Bayesian Information Criterion (BIC) as explained in Section SI 4.

\section{Comparison of Network- and Physical-Based Distances of Two Particles} 
\begin{figure}[!h]
    \centering
    \includegraphics[scale=0.5]{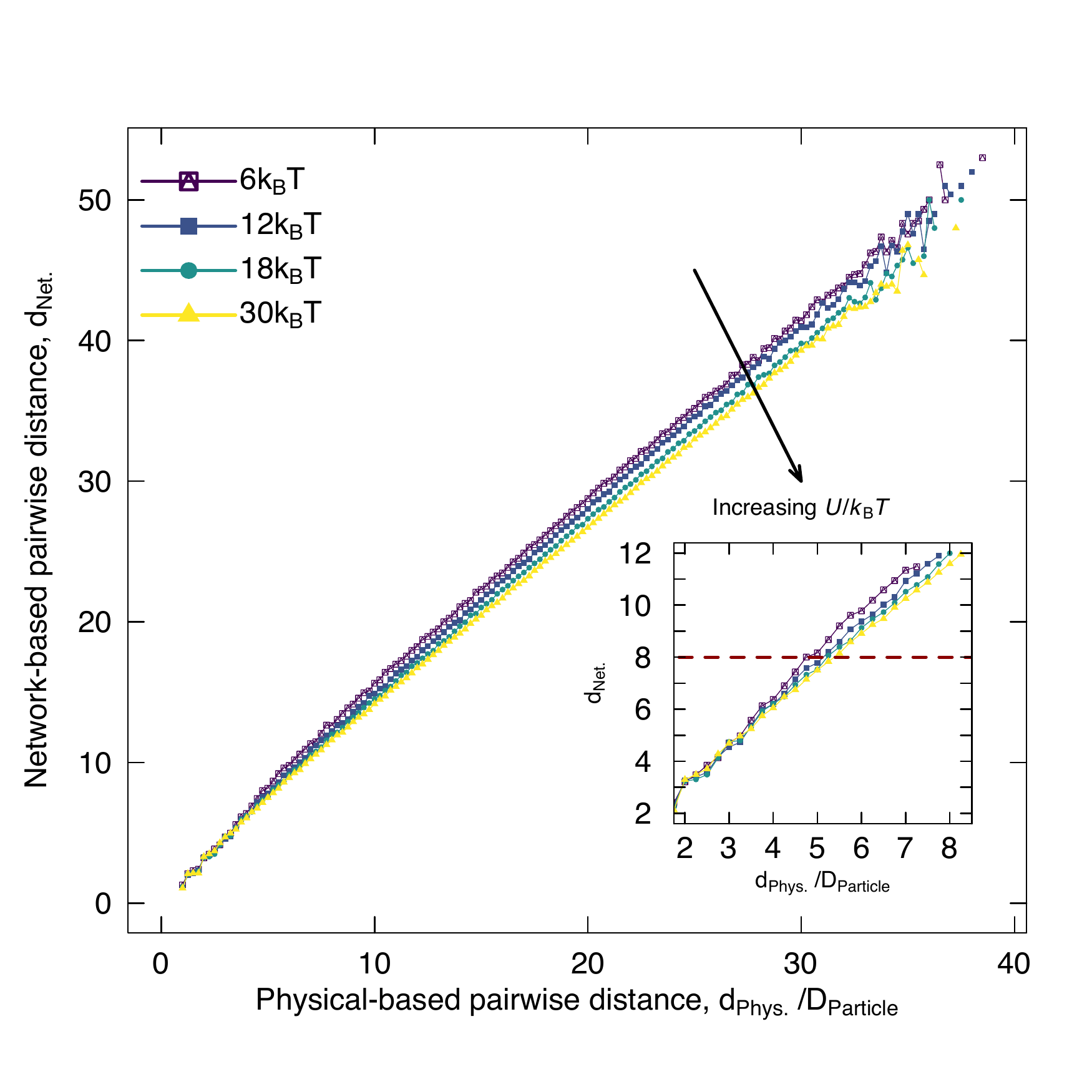}
    \caption{Network- and physical-based distances of particles in a gel network.}
    \label{fig_SI_physical_network_distance}
\end{figure}

\vspace{-20pt}
\noindent\rule{6.3in}{0.4pt}
The ability of a network to connect two particles placed in a given physical distance exhibits two distinct regimes of behavior. For every pair of particles in a gel network, we calculated their physical distance and the length of the shortest path connecting them in the network. Fig. \ref{fig_SI_physical_network_distance} shows that for particles connected by a path with eight edges or less, the average physical distance is a universal function of the network-based distance, while for nodes located farther apart, this correlation depends on the attraction strength.

\section{Bayesian Information Criterion (BIC)}
\begin{figure}[!h]
    \centering
    \includegraphics[trim={0 1cm 0 0},clip,scale=1.0]{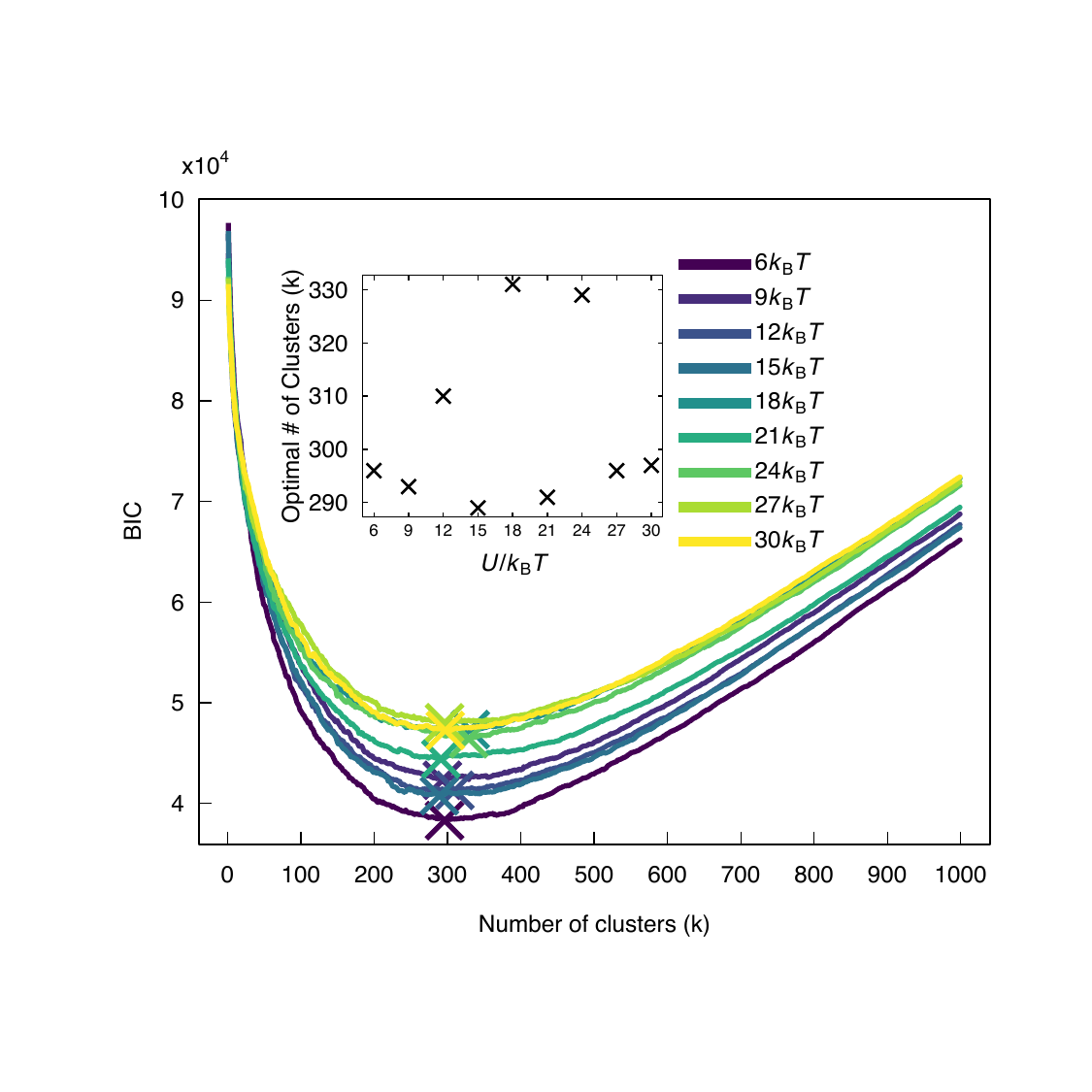}
    \caption{BIC versus number of clusters for all gel networks. The minimum is marked by a ($\times$) symbol for each attraction strength. The inset shows the selected value of $k$ for each gel.} 
    \label{SI_BIC}
\end{figure}

\vspace{-20pt}
\noindent\rule{6.3in}{0.4pt}
We chose the number of clusters ($k$) for the coarse-graining of a colloidal network into a cluster network by minimizing the BIC function. Fig. \ref{SI_BIC} shows the BIC values for a wide range of cluster numbers across all gel networks. The inset of Fig. \ref{SI_BIC} shows the optimum $k$ for each attraction strength used as an input to the GMM algorithm.

\blankpage
\section{Estimation of the bond stiffness of a chain} 
\begin{figure}[h]
    \centering
    \includegraphics{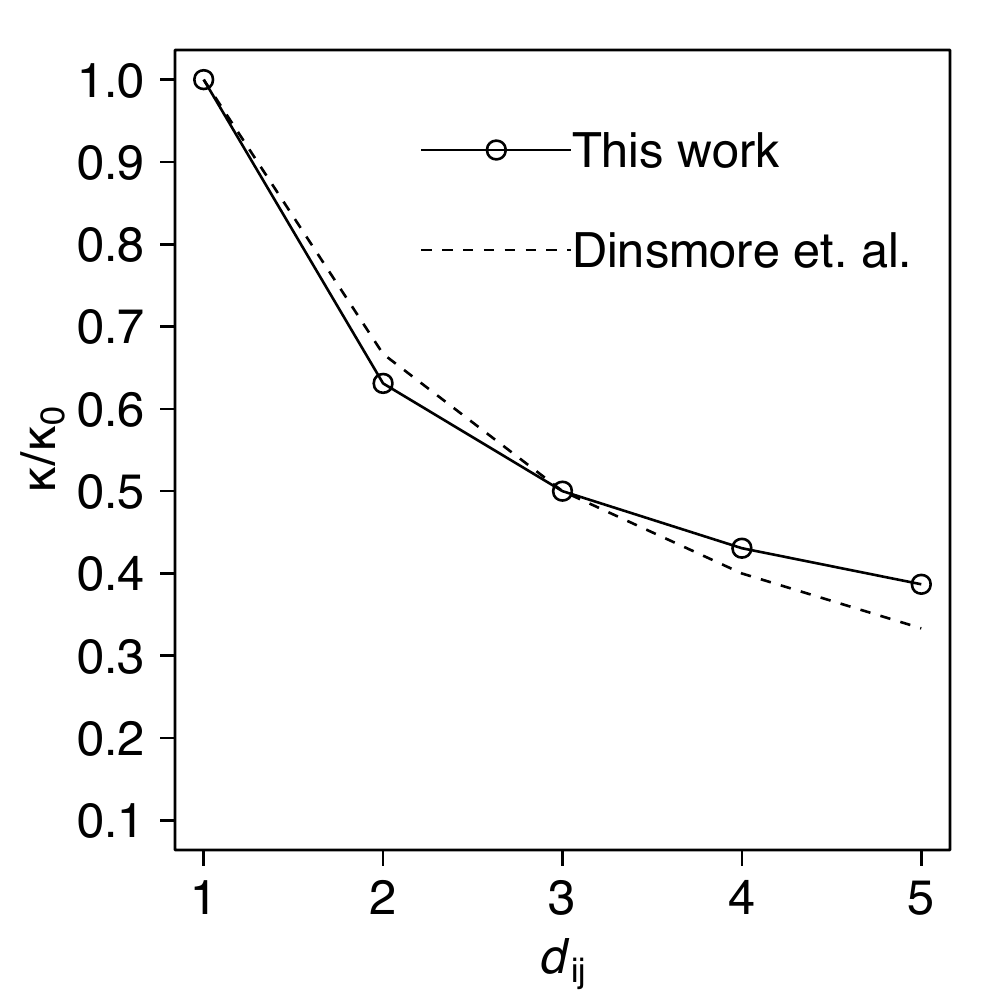}
    \caption{Bond stiffness calculation for the inter-cluster bonds, compared to experimentally measured values for a chain with the end-to-end distance of $d_{ij}$}
    \label{kappa_comp}
\end{figure}

\vspace{-20pt}
\noindent\rule{6.3in}{0.4pt}
For approximation of the stiffness of inter-cluster bonds used in the mass-spring calculations, we use the polymer chain stiffness approximation where the stiffness of a chain is approximated as a function of both the single particle-particle stiffness, and the size of the backbone of the chain. Specifically, we estimate the stiffness of a cluster-cluster connection of length $d_{ij}$ as $\textit{K}_s(\textit{U}/ \textit{k}_B \textit{T}, d_{ij}) = -\textit{K}_s(\textit{U}/ \textit{k}_B \textit{T},1) \times  log(\frac{1}{d_{ij}+1}))/log(1/2) $, where $\textit{K}_s(\textit{U}/ \textit{k}_B \textit{T},1)$ is the stiffness of a single particle-level bond and $d_{ij}$ is the lengths of the longest shortest path of the cluster (this is the equivalent of the size of the cluster backbone). The choice of such approximation is of course not unique but here we compare the stiffness obtained from this approximation, to the experimentally measured results of  Dinsmore et al. \cite{dinsmore2002direct}. They showed that the elastic modulus of a cluster whose end-to-end chain includes $\textit{N}$ particles is $\kappa(N) ~ \textit{N}^{-1}$.  In Fig \ref{kappa_comp}, our approximation of the bond stiffness is compared to the work of Dinsmore et al.\cite{dinsmore2002direct}.

\blankpage
\section{Correlation between Elasticity Loss and Edge Betweenness Centrality} 
\begin{figure}[!h]
    \centering
    \includegraphics[width=\textwidth]{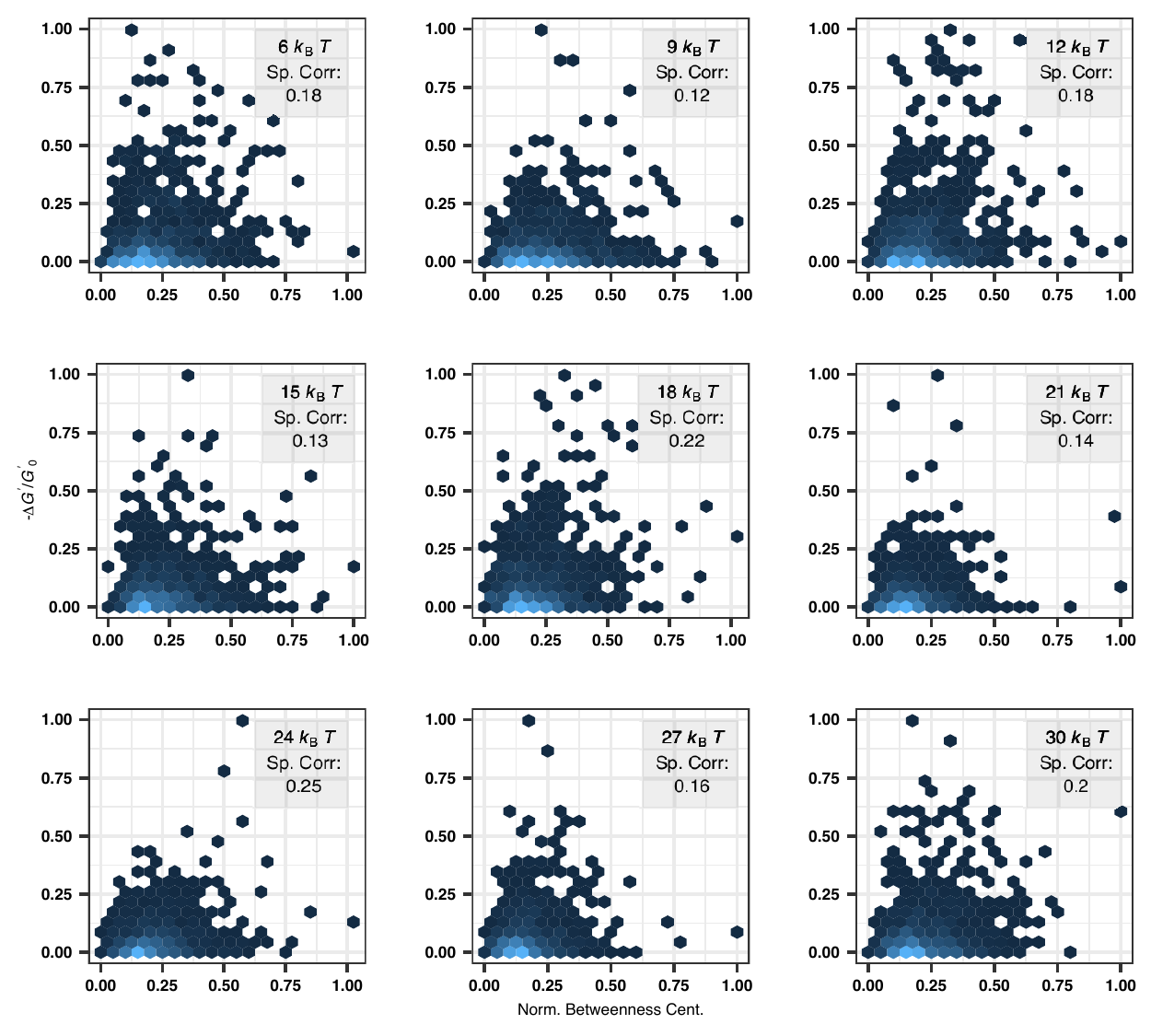}
    \caption{Correlation between elasticity-drop and betweenness centrality of an edge for all attraction strengths studied.} 
    \label{SI_betc_G_corr}
\end{figure}

\begin{figure}[!h]
    \centering
    \includegraphics[width=\textwidth]{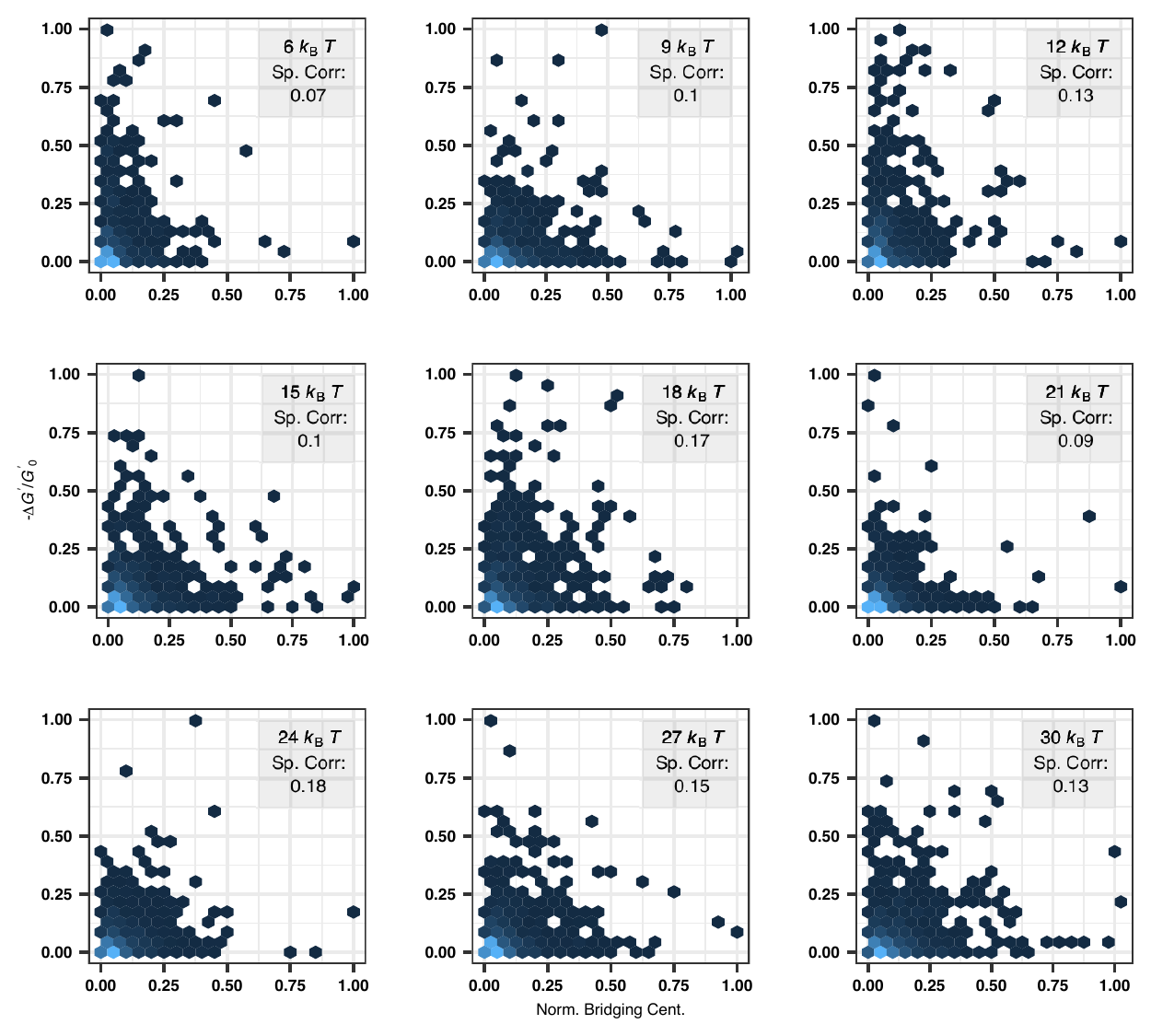}
    \caption{Correlation between elasticity drop and bridging centrality of an edge for all attraction strengths studied.}
    \label{SI_bridgeC_G_corr}
\end{figure}

\begin{figure}[!h]
    \centering
    \includegraphics[trim={4cm 6cm 4cm 6.cm},clip,width=\textwidth]{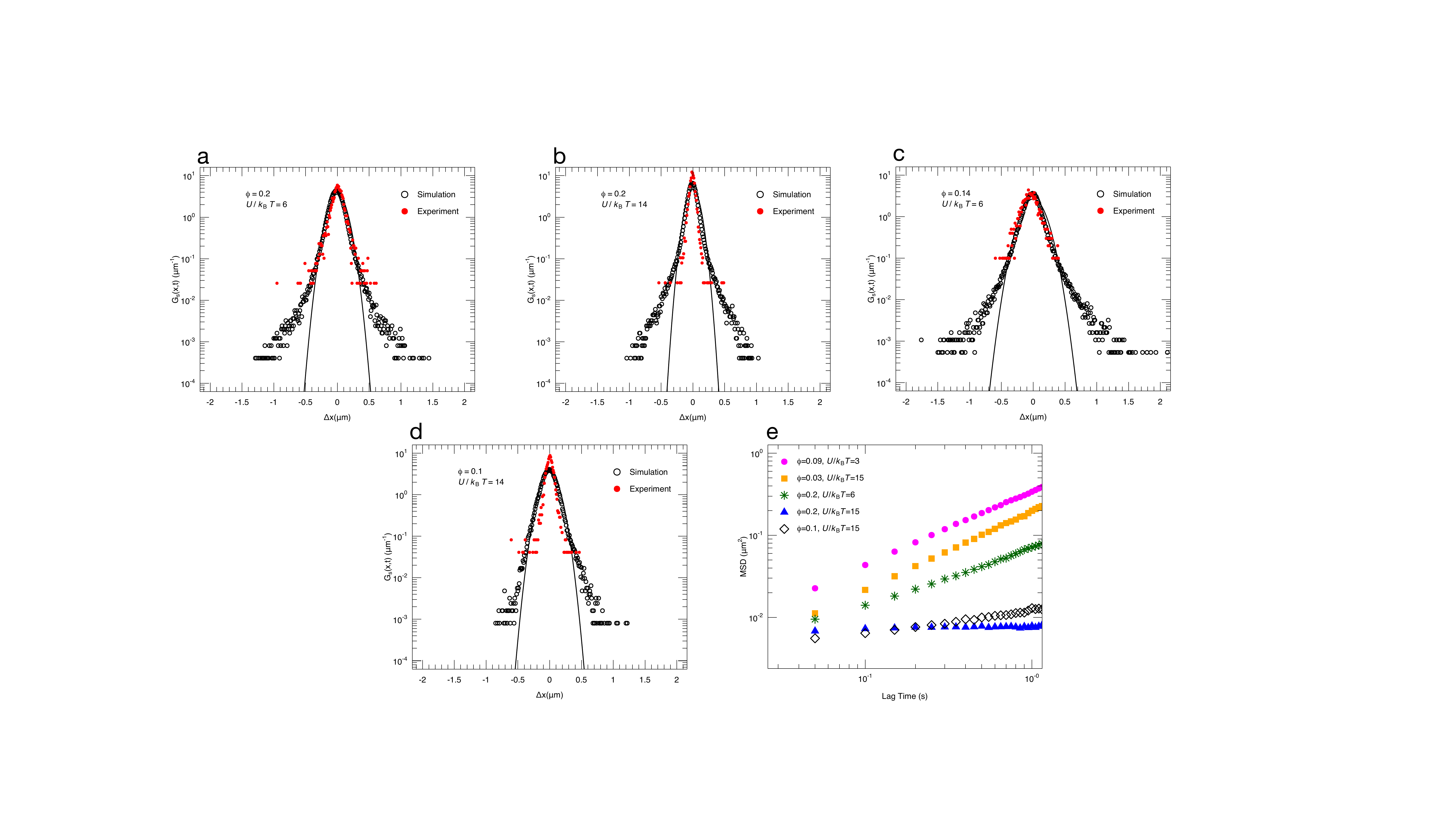}
    \caption{Van hove self-correlations of the particle displacement for  \textbf{(a)} $\phi=0.2$ and $6 \textit{k}_{\mathrm{B}} \textit{T}$, \textbf{(b)} $\phi=0.2$ and $14 \textit{k}_{\mathrm{B}} \textit{T}$, \textbf{(c)} $\phi=0.14$ and $6 \textit{k}_{\mathrm{B}} \textit{T}$, \textbf{(d)} $\phi=0.1$ and $14 \textit{k}_{\mathrm{B}} \textit{T}$, and \textbf{(e)} mean squared displacement (MSD) of particles obtained from confocal imaging.} 
    \label{Fig5}
\end{figure}

\vspace{-20pt}
\noindent\rule{6.3in}{0.4pt}
We applied Girvan-Newman algorithm to the coarse-grained network and calculated the relative change in elasticity after removing edges one by one from the networks. Given the results in Fig. \ref{SI_betc_G_corr}, there is a meaningful correlation between the amount of elasticity loss and the measure of betweenness centrality for each of the removed edges. Correlation between elasticity drop and bridging centrality of an edge are also plotted in Fig. \ref{SI_bridgeC_G_corr} for comparison.

\blankpage\blankpage
\section{van Hove self-correlation and mean squared displacement of particles} 
\vspace{-20pt}
\noindent\rule{6.3in}{0.4pt}
The van Hove self-correlation of particle motion, as well as mean squared displacement of the particle assembly are used as proof for the fluid vs. gel state of the strucuture. The arrested state of gel structure is visible in a quasi-plateau MSD curve as a function of lag time, as opposed to a diffusive or slightly sub-diffusive motion for the fluid state. Note that clusters of particles that do not construct a single system-spanning network show significantly hindered motion and thus are sub-diffusive; however those are not considered gelled as the structure evolved over time and further coarsens.

\blankpage

\end{document}